\documentclass[preprint]{rspublic}

\usepackage{graphicx}% Include figure files
\usepackage{dcolumn}% Align table columns on decimal point
\usepackage{bm}% bold math

\newcommand{\nab}{{\bf{\nabla}}}

\newcommand{\lam}{\Lambda(t)}
\newcommand{\ps}{\Psi_\alpha}
\newcommand{\pa}{\partial}

\newcommand{\largex}{{\bf x}}
\newcommand{\largep}{{\bf p}}

\newcommand{\largeB}{{\bf B}}
\newcommand{\largeA}{{\bf A}}

\newcommand{\ep}{\varepsilon}

\newcommand{\supp}{^{(0)}}
\newcommand{\supa}{^{(\alpha)}}
\newcommand{\largee}{{\bf e}}
\newcommand{\largeE}{{\bf E}}
\newcommand{\htheta}{\frac{\theta}{2}}

\begin{document}

%\preprint{Draft (version of February 25, 2008)}

\title{Fast-forward of quantum adiabatic dynamics in electro-magnetic field}% Force line breaks with \\

\author[Shumpei Masuda \& Katsuhiro Nakamura]{Shumpei Masuda$^{1}$ and Katsuhiro
 Nakamura$^{2,3}$}
%\email{syunpei710@yahoo.co.jp}
%\email{smasuda@amp.i.kyoto-u.ac.jp}
%\altaffiliation[Electric address:]{smasuda@}%Lines break automatically or can be forced with \\

%\author{Katsuhiro Nakamura$^{3,4}$}%
%\email{nakamura@a-phys.eng.osaka-cu.ac.jp}
\affiliation{%
$^{1}$
Department of Physics,
Kwansei Gakuin University,
Gakuen, Sanda, Hyogo 669-1337, Japan\\
%}%
%\begin{center}
%\affiliation{%
%$^2$
%Department of Applied Mathematics and Physics,
%Graduate School of Informatics, Kyoto University,
%Sakyo-ku, Kyoto 606-8501, Japan\\
%}%
%\affiliation{%
$^2$
Heat Physics Department,
Uzbek Academy of Sciences,
28 Katartal Str.,
100135 Tashkent,
Uzbekistan\\
$^3$
Department of Applied Physics,
Osaka City University,
Sumiyoshi-ku, Osaka 558-8585, Japan
}%

\date{\today}
\label{firstpage}
\maketitle

\begin{abstract}{mechanical control of atoms, electro-magnetic field,
spin dynamics}
We show a method to accelerate quantum adiabatic 
dynamics of wavefunctions under 
electro-magnetic field
by developing the previous theory (Masuda \& Nakamura 2008 and 2010).
Firstly we investigate the orbital dynamics of a charged particle.
We derive the driving field 
which accelerates quantum adiabatic dynamics
in order to obtain the final adiabatic states 
%We propose a method 
%to accelerate adiabatic dynamics of wavefunction in
%magnetic field to
%generate, in any desired short time, a final adiabatic state 
except for the spatially uniform phase such as the adiabatic phase 
in any desired short time.
Fast-forward of adiabatic squeezing and transport in the electro-magnetic 
field is exhibited.
%At the end of the acceleration the wavefunction becomes stationary again.
Secondly we investigate spin dynamics under the magnetic field,
showing the fast-forward of adiabatic spin inversion and of adiabatic dynamics
in Landau-Zener model.
The connection of the present framework 
with Kato-Berry's transitionless quantum driving is elucidated in Appendix.
\end{abstract}

%\pacs{81.16.Ta, 37.90.+j, 03.65.-w, 05.45.Yv}

\maketitle

\section{Introduction}
\label{Introduction}
The technology to manipulate tiny objects is rapidly evolving, and nowadays
we can control even individual atoms (Eigler \& Schweizer 1990).
And various methods to control quantum states have been reported 
in Bose Einstein condensates (BEC) (Leggett 2001; Ketterle 2002; Leanhardt 
{\it et al.} 2002; Gustavson $et. al.$ 2002), 
in quantum computing with the use of
spin states (Nielsen \& Chuang 2000)
%\cite{niel} 
and in many other fields of applied physics.
It would be important to consider the acceleration of such manipulations
of quantum states for manufacturing purposes and for innovation of technologies.
In our previous paper (Masuda \& Nakamura 2008), 
%\cite{masuda}
we proposed 
the acceleration of quantum dynamics with the use of the
additional phase of wavefunctions (WFs). 
We can accelerate a given quantum dynamics and exactly obtain 
a target state in any desired short time, where the target state is defined
as the final state in a given standard dynamics.
This kind of acceleration is called fast-forward of quantum dynamics.

%On the other hand, we also showed a method to accelerate the quantum 
%adiabatic dynamics \cite{masuda2}.
One of the most important application of the theory of fast-forward is the 
acceleration of quantum 
adiabatic dynamics (Masuda \& Nakamura 2010).
%\cite{masuda2}.
The adiabatic process occurs when the 
external parameter of Hamiltonian of the system is adiabatically changed.
Quantum adiabatic theorem (Born \& Fock 1928; Kato 1950; Messiah 1962)
%\cite{born,kato,mess}
states that, 
if the system is initially in an eigenstate of the instantaneous Hamiltonian, 
it remains so during the adiabatic process 
(Berry 1984; Aharonov \& Anandan 1987; Samuel \& Bhandari 1988;
Shapere \&  Wilczek 1989;
Nakamura \& Rice 1994; Bouwmeester $et. al.$ 1996;
 Farhi $et. al.$ 2001; Roland \& Cerf 2002;
 Sarandy \&  Lidar 2005).
%\cite{berry1,aha,samuel,shap,naka,bouw,farhi,roland,sara}.
%And it has been studied in various contexts 
The rate of change in the parameter of Hamiltonian with respect to time
is infinitesimal,
so that it takes infinite time to reach the final state in the adiabatic
process. 
However, by using our theory 
(Masuda \& Nakamura 2010),
%\cite{masuda2}, 
the target states (final adiabatic states) 
are available 
in any desired short time.
There, infinitesimally-slow change in the adiabatic dynamics 
is compensated by the infinitely-rapid
fast-forward.
% while the target state is originally 
%accessible after infinite time through the adiabatic dynamics.

%by using the reverse engineering approach \cite{kato}.
%(The connection between their theory and ours have not been 
%exactly clarified yet.)

On the other hand,
electro-magnetic field (EMF)
%\cite{thou,tsuru} 
is often used to control quantum states 
$e.g.$, 
in manifestation of the quantum Hall effect 
(Thouless $et. al.$ 1982)
%\cite{thou} 
and manipulation of 
Bose-Einstein Condensates 
(Ketterle 2002).
%\cite{Kett1}.
%Independently, 
%M. Berry showed the transitionless quantum driving of spin dynamics
%by applying the reverse engineering approach \cite{berry2}.
%The reverse engineering approach \cite{kato} gives the  
%Hamiltonian which generates the 
%adiabatic states in a finite time.
%In the general formalism, 
%however, we should consider all the transitions between 
%instantaneous eigenstates
%due to nonadiabatic change of the system.
The acceleration of the adiabatic dynamics in EMF is far from 
being trivial,
 and therefore it is highly desirable to extend our theory of the fast-forward
to systems under EMF.
In this paper, we extend our previous theory of the fast-forward 
to the system in EMF,
and derive a driving field to generate
the target adiabatic state exactly besides from a spatially uniform phase 
like dynamical and adiabatic phases (Berry 1984).
%\cite{berry1}. 

%After the acceleration, the WF becomes stationary again, and is in
%the same energy level as the initial one.
In section \ref{Regular fast-forward}, 
we explain the method of the standard fast-forwarding
under EMF. Section \ref{Fast-forward of adiabatic dynamics in magnetic field}
is devoted to the fast-forward of the adiabatic 
dynamics with EMF.
Examples of orbital dynamics of a charged particle
are given in Section \ref{Examples}.
Section \ref{Fast-forward in two-level systems in magnetic field} 
is concerned with spin dynamics in magnetic field (MF).
Summary and discussion are given in Section \ref{Conclusion}.
%In the last section, the fast-forward of spin dynamics in magnetic field (MF) 
%is shown. 
In Appendix, 
we shall elucidate the connection of the present framework with 
Kato-Berry's transitionless quantum driving. 

\section{Standard fast-forward}
\label{Regular fast-forward}
Before embarking on the fast-forward of adiabatic dynamics, 
we derive the driving electro-magnetic field (EMF) which
accelerates the (non-adiabatic) standard dynamics of wavefunction (WF)
and drive it from an initial state to the target state which is defined as
the final state of the standard dynamics. 
Hamiltonian for the system with electric field
$\largeE_0$ and MF $\largeB_0$ corresponding to vector potential 
$\largeA_0$ is written as
$H_0 = \frac{1}{2m_0}(\largep+\frac{e}{c}\largeA_0)^2$.
The electric and magnetic field are related to $\largeA_0$ as
$\largeE_0=-\frac{d\largeA_0}{dt}$ and $\largeB_0=\nab\times\largeA_0$,
respectively. 
For simplicity of notation, we put $\frac{e}{c}$ to be $1$ hereafter,
and we consider in the main text the case where
in $a \  priori$ there is no scalar potential $V_0$.
The results for systems with $V_0$ is shown in
\ref{Driving potential in systems with potential}.
The Schr$\ddot{\mbox{o}}$dinger equation with a nonlinearity constant
$c_0$ (appearing in macroscopic quantum dynamics) is represented as
\begin{eqnarray}
i\hbar\frac{\pa \Psi_0}{\pa t} &=& -\frac{\hbar^2}{2m_0}\nab^2\Psi_0
- \frac{i\hbar}{2m_0}(\nab\cdot\largeA_0)\Psi_0
- \frac{i\hbar}{m_0}\largeA_0\cdot\nab\Psi_0 
+ \frac{\largeA^2_0}{2m_0}\Psi_0\nonumber\\
&& -c_0|\Psi_0|^2\Psi_0.
\label{schr}
\end{eqnarray}
$\Psi_0$ is a known function of space $\largex$ and time $t$
and is called a standard state.
For any long time $T$ called a standard final time, we choose
$\Psi_0(t=T)$ as a target state that we are going to generate.
Let $\Psi_\alpha(\largex,t)$ be a virtually 
fast-forwarded state of $\Psi_0(\largex,t)$ defined by 
\begin{eqnarray}
|\Psi_\alpha(t)> = |\Psi_0(\alpha t)>,
\label{ps111}
\end{eqnarray}
where $\alpha$ ($>0$) 
is a time-independent magnification factor of the fast-forward.
%For $\alpha>1$, $\ps$ is a speeded state of $\Psi_0$ and 
%slowed one for $0<\alpha<1$ like a slow-motion of $\Psi_0$.
%A rewind can occur for $\alpha<0$,
%and the WF pauses when $\alpha=0$.

In general, the magnification factor can be time-dependent.
Hereafter $\alpha$ is assumed to be time-dependent, $\alpha = \alpha(t)$.
%The time-evolution of WF is accelerated and decelerated
%when $\alpha(t)$ is increasing and decreasing, respectively.
In this case,
the virtually fast-forwarded state is defined as,
\begin{eqnarray}
  |\Psi_\alpha(t)> = |\Psi_0(\lam)>,
\label{exact_s}
\end{eqnarray}
where
\begin{eqnarray}
\lam = \int_0^{t}\alpha(t')dt'.
\label{lam}
\end{eqnarray}

Since the generation of $\ps$ requires an anomalous mass reduction,
we can not generate $\ps$ as it stands (Masuda \& Nakamura 2008).
%\cite{masuda}. 
But we can obtain the target state by considering
a fast-forwarded state $\Psi_{FF} = \Psi_{FF}(\largex,t)$ which differs from
$\ps$ by an additional space-dependent phase,
$f=f(\largex, t)$, as
\begin{eqnarray}
\Psi_{FF}(t) = e^{if}\ps(t) = e^{if}\Psi_0(\lam),
\label{psiff0}
\end{eqnarray}
where $f=f(\largex,t)$ is a real function of space $\largex$ and time $t$ 
and is called the additional phase.
Schr$\ddot{\mbox{o}}$dinger equation for fast-forwarded state $\Psi_{FF}$
is supposed to be given as
\begin{eqnarray}
i\hbar\frac{\pa \Psi_{FF}}{\pa t} &=& -\frac{\hbar^2}{2m_0}\nab^2\Psi_{FF}
- \frac{i\hbar}{2m_0}(\nab\cdot\largeA_{FF})\Psi_{FF}
- \frac{i\hbar}{m_0}\largeA_{FF}\cdot\nab\Psi_{FF} 
+ \frac{\largeA^2_{FF}}{2m_0}\Psi_{FF}\nonumber\\
&& + V_{FF}\Psi_{FF} -c_0|\Psi_{FF}|^2\Psi_{FF}.
\label{schrff}
\end{eqnarray}
$V_{FF}=V_{FF}(\largex,t)$ and $\largeA_{FF}=\largeA_{FF}(\largex,t)$ 
are called a driving scalar potential and a driving vector 
potential, respectively.
Driving EMF is related with $V_{FF}$ and $\largeA_{FF}$ as 
\begin{eqnarray}
\largeE_{FF} &=& -\frac{d\largeA_{FF}}{dt} - \nab V_{FF},\label{eff}\\
\largeB_{FF} &=& \nab\times\largeA_{FF}.
\label{df1}
\end{eqnarray}
%We suppose $\Psi_{FF}$ is given by
%\begin{eqnarray}
%\Psi_{FF}(\largex,t) = e^{if}\Psi_0(\largex,\lam),
%\label{psiff0}
%\end{eqnarray}
%where $f=f(\largex,t)$ is a real function which is called additional phase,
%and $\lam$ is given by
%\begin{eqnarray}
%\lam = \int_0^{t}\alpha dt.
%\end{eqnarray}
%$\alpha$ is the magnification factor of the fast-forward which is real function
%of time and starts from $1$ and become 1 again at the final time of the 
%fast-forward $T_F$.
If we appropriately 
tune the time dependence of $\alpha$ (the detail will be shown later), 
the additional phase can vanish
at the final time of the fast-forward $T_{F}$, and we can obtain the 
exact target
state
\begin{eqnarray}
\Psi_{FF}(T_F) = \Psi_0(T),
\end{eqnarray}
where $T_F$ is the final time of the fast-forward defined by
\begin{eqnarray}
T=\int_0^{T_F}\alpha(t)dt.\label{tf1}
\end{eqnarray}

Equation (\ref{schrff}) is rewritten with the use of 
equations (\ref{schr}), (\ref{lam}) and 
(\ref{psiff0}) as
\begin{eqnarray}
-\hbar\frac{\pa f}{\pa t}\Psi_0 &+& \alpha[
-\frac{\hbar^2}{2m_0}\nab^2\Psi_0 - \frac{i\hbar}{2m_0}(\nab\cdot\largeA_0)\Psi_0
- \frac{i\hbar\largeA_0\cdot\nab\Psi_0}{m_0} + \frac{\largeA_0^2}{2m_0}\Psi_0
]\nonumber\\
&&= -\frac{\hbar^2}{2m_0}[\nab^2\Psi_0 + i(\nab^2 f)\Psi_0
+ 2i\nab f\cdot\nab\Psi_0 - (\nab f)^2\Psi_0] \nonumber\\
&& \ \ \ - 
\frac{i\hbar}{2m_0}(\nab\cdot\largeA_{FF})\Psi_0 - \frac{i\hbar}{m_0}\largeA_{FF}
\cdot\{ i(\nab f)\Psi_0 + \nab\Psi_0 \}
+ \frac{\largeA^2_{FF}}{2m_0}\Psi_0\nonumber\\ 
&& + V_{FF}\Psi_0 + (\alpha-1)c_0|\Psi_0|^2\Psi_0,
\label{eq1}
\end{eqnarray}
where $f(\largex,t)$, $\Psi_0(\largex,\lam)$, $\alpha(t)$, 
$\largeA_0(\largex,\lam)$, $\largeA_{FF}(\largex,t)$ and
$V_{FF}(\largex,t)$ are abbreviated by $f$, $\Psi_0$, $\alpha$, $\largeA_0$,
$\largeA_{FF}$ and $V_{FF}$, respectively.
The same abbreviation will be used throughout in this section.
Real and imaginary parts of equation (\ref{eq1}) divided 
by $\Psi_0$ yield a pair of equations as 
\begin{eqnarray} 
|\Psi_0|^2\nab\cdot(\nab f - \frac{\alpha\largeA_0-\largeA_{FF}}{\hbar})
&+&2\mbox{Re}[\Psi_0\nab\Psi_0^\ast](\nab f -
\frac{\alpha\largeA_0-\largeA_{FF}}{\hbar})\nonumber\\
&+&(\alpha-1)\mbox{Im}[\Psi_0\nab^2\Psi_0^\ast] = 0,
\label{f1}
\end{eqnarray}
and
\begin{eqnarray} 
\frac{V_{FF}}{\hbar} &=& 
-\frac{\pa f}{\pa t} - (\alpha-1)\frac{\hbar}{2m_0}\mbox{Re}
[\nab^2\Psi_0/\Psi_0]
\nonumber\\ 
&& -\frac{\hbar}{m_0} (\nab f 
- \frac{\alpha\largeA_0 - \largeA_{FF}}{\hbar})
\mbox{Im}[\nab\Psi_0/\Psi_0] - \frac{\hbar}{2m_0}(\nab f)^2
+ \frac{\hbar}{2m_0}\frac{\alpha\largeA_0^2-\largeA_{FF}^2}{\hbar^2}\nonumber\\
&& -\frac{\hbar}{m_0}\frac{\largeA_{FF}}{\hbar}\cdot\nab f 
-(\alpha-1)\frac{c_0}{\hbar}|\Psi_0|^2.
\label{vff1}
\end{eqnarray}
We can take the driving scalar potential from equation (\ref{vff1})
and the additional phase $f$ which is a solution of equation (\ref{f1}).

\subsection{Additional phase and driving field}
In order to obtain the driving field, we should first calculate the 
additional phase in equation (\ref{f1}).
Here we derive a general solution of equation (\ref{f1})
from the continuity equation for $\Psi_0$ and $\Psi_{FF}$.
With the use of equation (\ref{schr}), we have a continuity equation for $\Psi_0$
\begin{eqnarray}
\frac{\pa}{\pa t}|\Psi_0|^2
=\frac{\hbar}{m_0}\nab\cdot(\mbox{Im}[\Psi_0\nab\Psi_0^\ast]
- \frac{\largeA_0}{\hbar}|\Psi_0|^2 ),
\label{eq4}
\end{eqnarray}
 and by using equations (\ref{psiff0}) and (\ref{schrff}), the
continuity equation for $\Psi_{FF}$ is
\begin{eqnarray}
\frac{\pa}{\pa t}|\Psi_{FF}|^2
=
\frac{\hbar}{m_0}\nab\cdot(-\nab f|\Psi_0|^2 + \mbox{Im}[\Psi_0\nab\Psi_0^\ast]
- \frac{\largeA_{FF}}{\hbar}|\Psi_0|^2 ),
\label{eq5}
\end{eqnarray}
where $\Psi_{FF}(\largex,t)$ is abbreviated by $\Psi_{FF}$.
From equation (\ref{psiff0}) which is the definition of $\Psi_{FF}$, we have
a relation between time derivatives of $\Psi_0$ and $\Psi_{FF}$ as
\begin{eqnarray}
\frac{\pa}{\pa t}|\Psi_{FF}|^2 = \alpha\frac{\pa}{\pa t}|\Psi_0|^2.
\label{eq3}
\end{eqnarray}
Combining equations (\ref{eq4}), (\ref{eq5}) and (\ref{eq3}),
we have the gradient of the additional phase
\begin{eqnarray}
\nab f(\largex,t) = (\alpha-1)\mbox{Im}[\nab\Psi_0/\Psi_0]
+ (\alpha\frac{\largeA_0}{\hbar} - \frac{\largeA_{FF}}{\hbar}).
\label{nabf0}
\end{eqnarray}
Noting the equivalence of gauges for $\largeA_0$ and $\largeA_{FF}$ due to
the initial condition ($\largeA_0(t=0) = \largeA_{FF}(t=0)$), 
we can take the gradient of the additional phase and $\largeA_{FF}$ as
\begin{eqnarray}
\nab f &=& (\alpha-1)\mbox{Im}[\nab\Psi_0/\Psi_0]
,\label{nabf}\\
\largeA_{FF} &=& \alpha\largeA_0.\label{aff1}
\end{eqnarray}
And it is easily confirmed that equations (\ref{nabf}) and (\ref{aff1})
satisfy equation (\ref{f1}).
Equation (\ref{aff1}) implies that MF should be magnified by $\alpha$ times,
that is,  
\begin{eqnarray}
\largeB_{FF}(\largex,t) = \alpha(t)\largeB_0(\largex,\lam),
\label{bff1}
\end{eqnarray}
where $\largeB_0 = \nab\times\largeA_0$ and 
$\largeB_{FF}=\nab\times\largeA_{FF}$.
When the standard state $\Psi_0$ is written with its phase $\eta(\largex,t)$ as
\begin{eqnarray}
\Psi_0(\largex,t) = |\Psi_0(\largex,t)|e^{i\eta(\largex,t)},
\label{eta0}
\end{eqnarray}
equation (\ref{nabf}) leads the expression of the additional phase and its gradient as
\begin{subequations}
\begin{eqnarray}
\nab f(\largex,t) &=& (\alpha-1)\nab\eta(\largex,\lam),\label{nabf2}\\
f(\largex,t) &=& (\alpha-1)\eta(\largex,\lam).
\label{f5}
\end{eqnarray}
\label{f5_2}
\end{subequations}
In equation (\ref{f5}), a space-independent constant term was neglected.

%\subsection{Driving potential}
Substitution of equation (\ref{aff1}) into equation (\ref{vff1}) yields 
the driving scalar potential
\begin{eqnarray}
\frac{V_{FF}}{\hbar} &=& - \frac{\pa f}{\pa t}
 - (\alpha-1)\frac{\hbar}{2m_0}
\mbox{Re}[\nab^2\Psi_0/\Psi_0]\nonumber\\
&&-\frac{\hbar}{m_0}\nab f\cdot\mbox{Im}
[\nab\Psi_0/\Psi_0] -\frac{\hbar}{2m_0}(\nab f)^2
-\frac{\hbar}{m_0}\frac{\alpha\largeA_0}{\hbar}\cdot\nab f\nonumber\\
&&-\frac{\hbar}{2m_0}\alpha(\alpha-1)\frac{\largeA_0^2}{\hbar^2}-
(\alpha-1)\frac{c_0}{\hbar}|\Psi_0|^2,
\label{vff05}
\end{eqnarray}
where $f$ and $\nab f$ are given by equation (\ref{f5_2}).
Therefore, once we have $\Psi_0$ and $\largeA_0$,
the driving field can be obtained from equations (\ref{eff}) and 
(\ref{df1}) with the use of
equation (\ref{aff1}) and (\ref{vff05}).
By applying $\largeE_{FF}$ and $\largeB_{FF}$ against the initial standard state,
we can generate the target state
in any short time $T_F$ related to the standard final time  
through $\alpha$ by equation (\ref{tf1}). 

\section{Fast-forward of adiabatic dynamics}
\label{Fast-forward of adiabatic dynamics in magnetic field}
So far,
we presented the fast-forward of
the standard dynamics in EMF which enables to generate the target state in any
desired short time.
%However, since we must know the WF in standard dynamics at each time 
%to derive the driving potential, 
%in general, we need to calculate the standard dynamics $\Psi_0(\largex,t)$ by
%numerical integration of time-dependent Schr$\ddot{\mbox{o}}$dinger equation.
%And such calculation takes long time, 
%because usually the dynamics which we are going to 
%accelerate is slow and it takes long time to reach the target state.
%To overcome this difficulty, 
Here we show the fast-forward of adiabatic dynamics of WF under 
EMF in an analogous 
manner which we employed in the previous paper (Masuda \& Nakamura 2010).
%\cite{masuda2}. 
In the process, 
the regularized adiabatic dynamics is introduced as standard states and 
we fast-forward them with 
infinitely large magnification factor $\alpha$.
%Explanation for the regularization will be given later.
In the acceleration, the initial state is stationary and it becomes back to
stationary state at the end of the fast-forward.
%In the process it takes less work and time in calculation 
%to obtain the standard dynamics than the regular fast-forward shown in the 
%previous section, because in the adiabatic dynamics the WF
%is an energy eigenstate of the instantaneous Hamiltonian, 
%so we do not have to integrate 
%Schr$\ddot{\mbox{o}}$dinger equation 
%with respect to time to see the standard state.
However we can not directly apply the theory in last section 
to adiabatic dynamics,
because the stationary state is
just an energy eigenstate of the instantaneous Hamiltonian and, therefore, 
is not suitable to be fast-forwarded.
Thus we first need to regularize the adiabatic dynamics
for the fast-forward. 

Let us consider $\Psi_0$ under 
%Suppose there is a wave function $\Psi$ in a system $H$ with 
$\largeE_0$ and $\largeB_0$ corresponding to the 
vector potential $\largeA_0 = \largeA_0(\largex,R(t))$
which adiabatically varies,
where $R=R(t)$ is a parameter 
which is changed 
from constant $R_0$ as
\begin{eqnarray}
R(t) = R_0 + \ep t.
\label{R}
\end{eqnarray}
The constant value $\ep$ is the rate of adiabatic change in $R(t)$ 
with respect to time
and is infinitesimal, that is,
\begin{eqnarray}
\frac{d R(t)}{d t} &=& \varepsilon,\label{dRdt}\\
\ep &\ll& 1.
\end{eqnarray}
The EMF is changed through this parameter.
%For the simplicity, we suppose there is no scalar potential $V_0$ here.
%Thus, the intensity of electric field is $O(\ep)$ in the adiabatic dynamics.
Hamiltonian of the system is represented as
\begin{eqnarray}
H_0 = \frac{(\largep+\largeA_0(\largex,R(t)))^2}{2m_0},
\label{H}
\end{eqnarray}
and Schr$\ddot{\mbox{o}}$dinger equation for $\Psi_0$ is given as
\begin{eqnarray}
i\hbar\frac{\pa \Psi_0}{\pa t} = -\frac{\hbar^2}{2m_0}\nab^2\Psi_0
- \frac{i\hbar}{2m_0}(\nab\cdot\largeA_0)\Psi_0 
-\frac{i\hbar}{m_0}\largeA_0\cdot\nab\Psi_0
+\frac{\largeA_0^2}{2m_0}\Psi_0  - c_0|\Psi_0|^2\Psi_0,
\label{schr1}
\end{eqnarray}
where $c_0$ is a nonlinearity constant.
The results for systems with $V_0$ are shown in 
\ref{Driving potential in systems with potential}.
If a system is in the $n$-th energy eigenstate at the 
initial time, the adiabatic 
theorem guarantees that, in the limit $\ep\rightarrow 0$, 
$\Psi_0$ remains in the $n$-th energy eigenstate of 
the instantaneous Hamiltonian.
Then $\Psi_0$ is written as
\begin{eqnarray}
\Psi_0(\largex,t,R(t)) = \phi_n(\largex,R(t)) 
e^{-\frac{i}{\hbar}\int_0^t E_n(R(t'))dt'}
e^{i\Gamma(t)},
\label{adia1}
\end{eqnarray}
where $E_n=E_n(R)$ and 
$\phi_n=\phi_n(\largex,R)$ are the $n$-th energy eigenvalue and
eigenstate corresponding to the parameter
$R$, respectively,
and $\Gamma = \Gamma(t)$ is the adiabatic phase given by
\begin{eqnarray}
\Gamma(t) = i\int_{0}^t\int_{-\infty}^\infty d\largex dt
\phi_n^\ast\frac{d}{dt}\phi_n,
\label{gamma}
\end{eqnarray}
which is independent of space coordinates.
$\phi_n$ fulfills
\begin{eqnarray}
\frac{\pa \phi_n}{\pa t} &=& 0,\label{dphidt}\\
 -\frac{\hbar^2}{2m_0}\nab^2\phi_n
-\frac{i\hbar}{2m_0}(\nab\cdot\largeA_0)\phi_n
&-&\frac{i\hbar}{m_0}\largeA_0\cdot\nab\phi_n
+\frac{\largeA_0^2}{2m_0}\phi_n  -c_0|\phi_n|^2\phi_n= 
E_n\phi_n.\nonumber\\\label{tis}
\end{eqnarray}
The second and the third 
factors of the right hand side of equation (\ref{adia1})
are called dynamical and adiabatic phase factors, respectively,
which are also space-independent. (We will not intend to realize these phase in
the fast-forward.)
The adiabatic dynamics in the 
limit $\ep\rightarrow 0$ takes infinitely long time
until we obtain an aimed adiabatic state (target state).

%During the fast-forward, we will take the limit $\ep\rightarrow 0$, 
%$\alpha\rightarrow\infty$, $T\rightarrow\infty$ so that $\alpha\ep \sim 1$.
%In this acceleration, we do not care about spatially uniform phase.
%(The spatially uniform phase can be controlled by spatially uniform
%potential, if it is necessary.)
%\footnote{The spatially uniform phase can be control by spatially uniform
%potential, if it is necessary.}.

For the fast-forward of the adiabatic dynamics, 
we should first choose an appropriate standard state and Hamiltonian.
%As mentioned in (Masuda \& Nakamura 2010),
%\cite{masuda2},
The original adiabatic state is not appropriate as the standard state.
%By following our previous work \cite{masuda2}, 
%But there is some ambiguity in the choice of the standard state and 
%Hamiltonian,
%because the state which we want to accelerate is represented in the limit
%$\ep\rightarrow 0$ and we will fast-forward it with infinitely-large 
%magnification factor. 
%One might think that we can take $\Psi_0$ and $H_0$ 
%in Eqs.(\ref{adia1}) and (\ref{H}) as a standard state and Hamiltonian.
%However, such idea is not proper
%because the state in equation (\ref{adia1}) is in an expression of WF 
%in the limit $\ep\rightarrow 0$ and does not 
%satisfy Schr$\ddot{\mbox{o}}$dinger equation
%up to $O(\ep)$ with small but finite $\ep$ \cite{kato,wu}. 
%In other words quantum dynamics in equation (\ref{schr1}) with finite $\ep$
%inevitably induces nonadiabatic transition, but $\Psi_0$ in
%equation (\ref{adia1})
%ignores such transitions.
%But the transition radically affect to the acceleration with infinitely
%large $\alpha$, and make the fast-forward fail.
Quantum dynamics in equation (\ref{schr1}) with small but finite $\ep$
inevitably induces non-adiabatic transition, but $\Psi_0$ in 
equation (\ref{adia1}) ignores such transition.
To overcome this difficulty,
we regularize the
standard state and Hamiltonian corresponding to
the adiabatic dynamics (Masuda \& Nakamura 2010),
%\cite{masuda2}, 
so that the following two conditions are satisfied.
% which agree with $\Psi$ in equation (\ref{adia1})
%and $V(\largex,R(t))$,respectively, 
%and satisfy Schr$\ddot{\mbox{o}}$dinger equation
%up to $O(\ep)$. 

%In the fast-forward of the adiabatic dynamics
%with the limit $\ep\rightarrow 0$, 
%$\alpha\rightarrow\infty$ and $\alpha\ep \sim 1$,
%the standard final time $T$ is chosen as $T = O(\frac{1}{\ep})$.
%In this case, a regularized standard state and Hamiltonian
%should fulfill the following conditions:\\
$1$. A regularized 
standard Hamiltonian and state of the fast-forward should agree
with  $H_0$ and $\Psi_0$ 
except for 
space-independent phase, respectively, in the limit $\ep\rightarrow 0$;

$2$. The regularized standard state should satisfy 
the Schr$\ddot{\mbox{o}}$dinger equation corresponding 
to the regularized standard Hamiltonian up to $O(\ep)$ with small but 
finite $\ep$.
%Because the form of the driving scalar 
%potential in equation (\ref{vff1}) is guaranteed 
%under this condition.
%However, both $H_0$ in equation (\ref{H}) and $\Psi_0$ in equation (\ref{adia1})
%does not satisfy the above conditions.

Hereafter $\Psi_0^{(reg)}$ and $H_0^{(reg)}$ denote
the regularized standard state and Hamiltonian, respectively, which
fulfill the conditions $1$ and $2$.
In the fast-forward, we take the limit
$\ep\rightarrow 0$, 
$\alpha\rightarrow\infty$ and $\alpha\ep \sim 1$.
Applying this regularization procedure in advance, the adiabatic dynamics
$\phi_n(R(0))\rightarrow \phi_n(R(T))$ can be accelerated and 
%aside from the dynamical and adiabatic phases, and obtain
the target state $\phi_n(R(T))$ is realized in any desired short time, 
where $T$ is a standard final time which is taken to be $O(1/\ep)$.

\subsection{Regularization of standard state}
Let us regularize the standard state so that it can be fast-forwarded
with infinitely large magnification factor.
Let us consider a regularized Hamiltonian $H_0^{(reg)}$
\begin{eqnarray}
H_0^{(reg)} = \frac{(\largep+\largeA^{(reg)})^2}{2m_0} + V_0^{(reg)}.
\label{H0}
\end{eqnarray}
The scalar potential $V_0^{(reg)}$ in the regularized
Hamiltonian
is given as
\begin{eqnarray}
V_0^{(reg)}(\largex,t) = \varepsilon \tilde{V}(\largex,t).
\label{sp1}
\end{eqnarray}
$\tilde{V}$ is a real function of $\largex$ and $t$ to be determined 
{\it a posteriori}, which is introduced to incorporate the effect of 
non-adiabatic transitions.
On the other hand, for the vector potential we put 
$\largeA^{(reg)}(\largex,t)=\largeA_0(\largex,t)$.
It is obvious that 
$H_0^{(reg)}$ agrees with $H_0$ in the limit $\ep \rightarrow 0$, that is,
\begin{eqnarray}
\lim_{\ep\rightarrow 0}H_0^{(reg)}(\largex,t) = H_0(\largex,R(t)).
\end{eqnarray}

%Suppose the magnetic field (MF) and the potential is adiabatically changed 
%with time,
%\begin{eqnarray}
%\largeA_0&=&\largeA_0(\largex,R(t)),\\
%V_0&=&V_0(\largex,R(t)),
%\end{eqnarray}
%where 
%\begin{eqnarray}
%R(t) = R_0 + \ep t, 
%\end{eqnarray}
%and $\ep$ is infinitensimal and $R_0$ is constant.

The standard state in the adiabatic dynamics should fulfill
Schr$\ddot{\mbox{o}}$dinger equation up to $O(\ep)$.
We suppose that a regularized standard state is given by
\begin{eqnarray}
\Psi_0^{(reg)} = \phi_ne^{-\frac{i}{\hbar}\int_0^{t}E_n(R(t'))dt'}e^{i\ep\theta},
\label{reg1}
\end{eqnarray}
where $\theta = \theta(\largex,t)$ is real, and $\phi_n=\phi_n(\largex,R(t))$ 
and $E_n=E_n(R(t))$ are the $n$-th energy eigenstate without dynamical phase
factor and eigenvalue of the original Hamiltonian $H_0$.
$\phi_n$ satisfies the instantaneous eigenvalue problem in equation (\ref{tis}).
%Schr$\ddot{\mbox{o}}$dinger equation
%\begin{eqnarray}
%E_n\phi_n = -\frac{\hbar^2}{2m_0}\nab^2\phi_n
%-\frac{i\hbar}{2m_0}(\nab\cdot\largeA_0)\phi_n
%-\frac{i\hbar}{m_0}\largeA_0\cdot\nab\phi_n
%+\frac{\largeA_0^2}{2m_0}\phi_n - c_0|\phi_n|^2\phi_n.
%\label{eig1}
%\end{eqnarray}
The Schr$\ddot{\mbox{o}}$dinger equation for regularized standard system 
is represented as
\begin{eqnarray}
i\hbar\frac{\pa \Psi_0^{(reg)}}{\pa t} &=& 
-\frac{\hbar^2}{2m_0}\nab^2\Psi_0^{(reg)}
- \frac{i\hbar}{2m_0}(\nab\cdot\largeA_0)\Psi_0^{(reg)} -
\frac{i\hbar}{m_0}\largeA_0\cdot\nab\Psi_0^{(reg)}
+\frac{\largeA_0^2}{2m_0}\Psi_0^{(reg)}\nonumber\\ 
&& + \ep\tilde{V}\Psi_0^{(reg)} -c_0|\Psi_0^{(reg)}|^2
\Psi_0^{(reg)}.
\label{schr2}
\end{eqnarray}
Substituting equation (\ref{reg1}) into equation (\ref{schr2}) 
and eliminating the equation of $O(1)$ with the use of equation (\ref{tis}),
we find the equation for $O(\ep)$:
%\begin{eqnarray}
%i\hbar\frac{\pa\phi_n}{\pa R}\ep + E_n\phi_n - 
%\hbar\frac{\pa\theta}{\pa t}\ep\phi_n &=& 
%-\frac{\hbar^2}{2m_0}[\nab^2\phi_n + 2i\ep\nab\theta\cdot\nab\phi_n
%-\ep^2(\nab\theta)^2\phi_n + i\ep(\nab^2\theta)\phi_n]\nonumber\\
%&& + \ep\tilde{V}\phi_n
%-\frac{i\hbar}{2m_0}(\nab\cdot\largeA_0)\phi_n\nonumber\\
%&&-\frac{i\hbar}{m_0}\largeA_0\cdot\nab\phi_n
%- \frac{i\hbar}{m_0}\largeA_0(i\ep\nab\theta)\phi_n
%+\frac{\largeA_0^2}{2m_0}\phi_n - c_0|\phi_n|^2\phi_n.
%\label{eq6}
%\end{eqnarray}
%By utilizing equation (\ref{eig1}) in equation (\ref{eq6})
\begin{eqnarray}
i\hbar\frac{\pa\phi_n}{\pa R} - \hbar\frac{d\theta}{d t}\phi_n
= -\frac{\hbar^2}{2m_0}[2i\nab\theta\cdot\nab\phi_n + 
i(\nab^2\theta)\phi_n] + \tilde{V}\phi_n
+\frac{\hbar}{m_0}\largeA_0\cdot(\nab\theta)\phi_n.\nonumber\\
\label{eq7}
\end{eqnarray}
Multiplying equation (\ref{eq7}) by $\frac{i}{\hbar}\phi_n^\ast$ 
and taking the
%\begin{eqnarray}
%-\phi_n^\ast\frac{\pa \phi_n}{\pa R} - i\frac{d\theta}{dt}|\phi_n|^2
%=\frac{\hbar}{2m_0}[2\phi_n^\ast\nab\phi_n\cdot\nab\theta
%+ |\phi_n|^2\nab^2\theta]
%+\frac{i}{\hbar}\tilde{V}|\phi_n|^2
%+\frac{i\hbar}{m_0}\frac{\largeA_0}{\hbar}\cdot\nab\theta|\phi_n|^2.
%\label{eq8}
%\end{eqnarray}
the real and imaginary parts of the resultant equation, we have
\begin{eqnarray}
|\phi_n|^2\nab^2\theta + 2\mbox{Re}[\phi_n^\ast\nab\phi_n]\cdot\nab\theta
+\frac{2m_0}{\hbar}\mbox{Re}[\phi_n^\ast\frac{\pa\phi_n}{\pa R}] = 0,
\label{theta1}
\end{eqnarray}
\begin{eqnarray}
\frac{\tilde{V}}{\hbar} = -\mbox{Im}[\frac{\pa\phi_n}{\pa R}/\phi_n]
%-\frac{d\theta}{dt}
-\frac{\hbar}{m_0}\mbox{Im}[\nab\phi_n/\phi_n]\cdot\nab\theta
-\frac{\hbar}{m_0}[\frac{\largeA_0}{\hbar}\cdot\nab\theta].
\label{vtilde}
\end{eqnarray}
From equation (\ref{theta1}), $\theta$ turns out to be 
dependent on $t$ only through $R(t)$.
Therefore the minor term $d\theta/dt (=\ep\frac{\pa\theta}{\pa R})$ was
suppressed in Eq(\ref{vtilde}).
%can be taken not to be a function of $t$ 
%explicitly. Then $\frac{d\theta}{dt} = \ep\frac{\pa \theta}{\pa R}$.
%The second term on the right hand side 
%in equation (\ref{vtilde0}) can be ignored because it concerns with 
%$O(\ep^2)$ of the potential. Therefore we have
%\begin{eqnarray}
%\frac{\tilde{V}}{\hbar} = -\mbox{Im}[\frac{\pa\phi_n}{\pa R}/\phi_n]
%-\frac{\hbar}{m_0}\mbox{Im}[\nab\phi_n/\phi_n]\cdot\nab\theta
%-\frac{\hbar}{m_0}[\frac{\largeA_0}{\hbar}\cdot\nab\theta].
%\label{vtilde}
%\end{eqnarray}
Equations (\ref{theta1}) and (\ref{vtilde}) give $\theta$ and $\tilde{V}$, 
respectively.
It is worth noting that $\theta$ is not 
explicitly affected by EMF. 

\subsection{Additional phase and driving field for fast-forward of 
adiabatic dynamics}
The regularized standard state in equation (\ref{reg1}) is now written as
\begin{eqnarray}
\Psi^{(reg)}_0 = |\phi_n|e^{i(\eta+\ep\theta)}e^{-\frac{i}{\hbar}\int_0^tE_n dt},
\label{reg2}
\end{eqnarray}
where $\eta$ is defined as a phase of $\phi_n$ by $\phi_n=|\phi_n|e^{i\eta}$.
By using $\Psi_0^{(reg)}$ in equation 
(\ref{reg2}) instead of  $\Psi_0$ in equation (\ref{f1}), we have
\begin{eqnarray}
&&|\phi_n|^2\nab\cdot(\nab f - \frac{\alpha\largeA_0-\largeA_{FF}}{\hbar})
+ 2[|\phi_n|\nab|\phi_n|](\nab f-\frac{\alpha\largeA_0-\largeA_{FF}}{\hbar})
\nonumber\\
&&-(\alpha-1)[2|\phi_n|\nab(\eta+\ep\theta)\cdot\nab|\phi_n|
+|\phi_n|^2\nab^2(\eta+\ep\theta)] = 0,
\label{eq10}
\end{eqnarray}
where $\phi_n(\largex,R(\lam))$, $f(\largex,t)$, $\largeA_0(\largex,R(\lam))$,
$\largeA_{FF}(\largex,t)$, $\eta(\largex,R(\lam))$ and 
$\theta(\largex,R(\lam))$ 
are
abbreviated by $\phi_n$, $f$, $\largeA_0$,
$\largeA_{FF}$, $\eta$ and $\theta$ respectively, 
and the same abbreviations will be taken hereafter in this section. 

Multiplying $\phi_n^\ast$ on both sides of equation (\ref{tis}) and taking 
its imaginary part with the use of $\phi_n=|\phi_n|e^{i\eta}$, we have
\begin{eqnarray}
\frac{\hbar}{m_0}[2|\phi_n|\nab\eta\cdot\nab|\phi_n|
+\nab^2\eta|\phi_n|^2] + \frac{\hbar}{m_0}[\nab\cdot\frac{\largeA_0}{\hbar}|\phi_n|^2
+ 2|\phi_n|\nab|\phi_n|\cdot\frac{\largeA_0}{\hbar}]
= 0.\nonumber\\
\label{eq9}
\end{eqnarray}
With the use of equation (\ref{eq9}) in equation (\ref{eq10}), we obtain
\begin{eqnarray}
|\phi_n|^2\nab\cdot(\nab f-\frac{\largeA_0-\largeA_{FF}}{\hbar})
+ 2[|\phi_n|\nab|\phi_n|](\nab f-\frac{\largeA_0-\largeA_{FF}}{\hbar})
\nonumber\\
- (\alpha-1)\ep[2|\phi_n|\nab\theta\cdot\nab|\phi_n| + 
|\phi_n|^2\nab^2\theta] = 0.
\label{f2}
\end{eqnarray}
We can easily confirm that 
\begin{eqnarray}
\nab f-\frac{\largeA_0-\largeA_{FF}}{\hbar} = (\alpha-1)\ep\nab\theta
\label{f3}
\end{eqnarray}
satisfies equation (\ref{f2}).
Noting $\largeA_{FF}(t=0) = \largeA_0(t=0)$,
we have the vector potential $\largeA_{FF}$ 
and gradient of the additional phase from equation (\ref{f3}) as
\begin{eqnarray}
\largeA_{FF}(t)=\largeA_0(\lam)\label{aff2}\\
\nab f = (\alpha-1)\ep\nab\theta,
\label{f4}
\end{eqnarray}
which should be compared with the result in equations (\ref{dphidt})
and (\ref{tis}) in the case of the standard fast-forward.
It is noteworthy that we do not have to magnify the MF
for the fast-forward, while in the standard fast-forward 
we need to magnify the MF 
by $\alpha$ times as shown in section
\ref{Regular fast-forward}.
The result in equation (\ref{f3}) is also obtained from the continuity equation.

%\subsection{Driving potential}
As mentioned in Section \ref{Regular fast-forward}, 
in the standard fast-forward with a standard scalar potential $V_0$,
the driving scalar potential is given by equation (\ref{eqa1}).
By using $\Psi_0^{(reg)}$ in equation (\ref{reg1}) and $V_0^{(reg)}$ 
in equation (\ref{sp1}) instead of $\Psi_0$ and $V_0$ and noting 
equations (\ref{tis}), (\ref{vtilde}), (\ref{aff2}) and (\ref{f4}),
equation (\ref{eqa1}) leads to 
the driving scalar potential as 
%By substituting equation (\ref{reg1}) into equation (\ref{vff1}) the driving potential is 
%given as
%\begin{eqnarray}
%\frac{V_{FF}}{\hbar} &=& -\frac{df}{dt} 
%-(\alpha-1)\frac{\hbar}{2m_0}\{ \mbox{Re}[\frac{\nab^2\phi_n}{\phi_n}]
%-2\ep\nab\theta\cdot\mbox{Im}[\frac{\nab\phi_n}{\phi_n}] -\ep^2(\nab\theta)^2\}
%\nonumber\\
%&&-\frac{\hbar}{m_0}(\nab f-\frac{\alpha\largeA_0-\largeA_{FF}}{\hbar})
%\{\mbox{Im}[\frac{\nab\phi_n}{\phi_n}]+\ep\nab\theta \}
%+\frac{\hbar}{m_0}\frac{\largeA_{FF}}{\hbar}\cdot\nab f
%\nonumber\\
%&&-\frac{\hbar}{2m_0}(\nab f)^2 + \frac{\hbar}{2m_0}
%\frac{\alpha\largeA_0^2-\largeA_{FF}^2}{\hbar^2}
%-(\alpha-1)\frac{c_0}{\hbar}|\phi_n|^2 + \alpha\ep\frac{\tilde{V}}{\hbar}.
%\label{vff2}
%\end{eqnarray}
%Multiplying equation (\ref{eig1}) by $\frac{1}{\phi_n\hbar}$ and taking its
%real part,  
%\begin{eqnarray}
%-\frac{\hbar}{2m_0}\mbox{Re}[\frac{\nab^2\phi_n}{\phi_n}]+\frac{\hbar}{m_0}
%\frac{\largeA_0}{\hbar}\cdot\mbox{Im}[\frac{\nab\phi_n}{\phi_n}]
%+\frac{\hbar}{2m_0}\frac{\largeA^2}{\hbar^2}+\frac{V_0}{\hbar}
%-\frac{c_0}{\hbar}|\phi_n|^2 = \frac{E_n}{\hbar},\label{eqeigen0}
%\end{eqnarray}
%where $E_n(R(\lam))$ is abbreviated by $E_n$.
%With the use of Eqs.(\ref{tis}), (\ref{aff2}) and (\ref{f4}) 
% in equation (\ref{vff2}),
%othe driving potential is represented as
\begin{eqnarray}
\frac{V_{FF}}{\hbar} &=& (\alpha-1)\frac{E_n}{\hbar}
-\frac{d\alpha}{dt}\ep\theta-\alpha^2\ep^2\frac{\pa\theta}{\pa R}
-\frac{\hbar}{2m_0}\alpha^2\ep^2(\nab\theta)^2
-\frac{\hbar}{m_0}\alpha\ep\frac{\largeA_0}{\hbar}\cdot\nab\theta
\nonumber\\
&&-\alpha\ep\mbox{Im}[\frac{\pa\phi_n}{\pa R}/\phi_n]
-\alpha\ep\frac{\hbar}{m_0}
\mbox{Im}[\frac{\nab\phi_n}{\phi_n}]\cdot\nab\theta,
\end{eqnarray}
where we omitted a term of $O(\ep)$.
While the first term diverges with infinitely large $\alpha$, 
it concerns only with spatially uniform phase of WF, which
we do not care about in the fast-forward and can be omitted.
%because it is spatially uniform.
%Since we do not care about the spatially uniform phase in the fast-forward,
%we omit this term 
Consequently, we
have the driving scalar potential
\begin{eqnarray}
\frac{V_{FF}}{\hbar} &=& 
-\frac{d\alpha}{dt}\ep\theta-\alpha^2\ep^2\frac{\pa\theta}{\pa R}
-\frac{\hbar}{2m_0}\alpha^2\ep^2(\nab\theta)^2
-\frac{\hbar}{m_0}\alpha\ep\frac{\largeA_0}{\hbar}\cdot\nab\theta
\nonumber\\
&&-\alpha\ep\mbox{Im}[\frac{\pa\phi_n}{\pa R}/\phi_n]
-\alpha\ep\frac{\hbar}{m_0}
\mbox{Im}[\frac{\nab\phi_n}{\phi_n}]\cdot\nab\theta.
\label{vff4}
\end{eqnarray}
The driving field can be obtained from 
equations (\ref{df1}), (\ref{aff2}) and 
(\ref{vff4}).
So far we considered the fast-forward in the systems without 
scalar potential $V_0$.
The driving scalar potential for systems with $V_0$ is shown in 
\ref{Driving potential in systems with potential}, while the driving vector
potential has the same form as in systems without $V_0$. 

The present theory of the fast-forward is different from
the reverse engineering approach based on the inverse technique
($e.g.$, Palao $et. al.$ 1998)
in the following sense: 
the latter approach is concerned with a direct fast-forward of the adiabatic 
state itself.
By contrast,
we consider the adiabatic states except for the spatially uniform phase but 
together with a controllable additional phase, and combine an idea of 
the infinitely-fast-forward and infinitesimally-slow adiabatic dynamics.
Muga $et. al.$ employed the reverse engineering approach to accelerate the 
adiabatic squeezing or expanding of BEC wave packet  
(Muga $et. al.$ 2009; Chen $et. al.$ 2010)
in a tunable harmonic trap,
finding a promising time dependence of the trapping frequency.
While their approach is limited to WF under the harmonic trap,
the present theory of the fast-forward enables to accelerate adiabatic 
dynamics of WF in any potential and EMF.

\section{Examples}
\label{Examples}
So far we showed theoretical framework of the fast-forward of 
adiabatic dynamics in EMF.
Here we give some examples of the fast-forward of adiabatic dynamics 
without nonlinearity constant ($c_0=0$) for simplicity. 
Our purpose is the realization
of target states defined in the adiabatic process 
in any desired short time,
while the target states are reached through infinitely long time in the 
original adiabatic dynamics.
In the following examples,
the magnification factor is commonly chosen (for $0\le t \le T_F$) in the form
\begin{eqnarray}
%\alpha(t)\ep = \bar{v}\cos(\frac{2\pi}{T_F}t + \pi) + \bar{v},
\alpha(t)\ep = \bar{v}(1-\cos(\frac{2\pi}{T_F}t)),
\label{valpha1}
\end{eqnarray}
where $\bar{v}$ is time average of $\alpha(t)\ep$ during the fast-forwarding, 
and the final time of the fast-forward
$T_F$ is related to the standard final time $T$ with $\bar{v}$ as 
$T_F=\ep T/\bar{v}$ (see equation (\ref{tf1})).
$\ep T$ and $T_F$ are taken as any finite value, 
although $\ep$ is infinitesimal and $T$ is
infinitely-large. Namely we aim to generate the target state in finite time,
while the state is supposed to be obtained after infinitely long time $T$
in the original adiabatic dynamics. 
%The time dependence of $\ep\alpha$ is shown for $0\le t\le T_{F}$ in 
%figure \ref{alpha}. 
$\alpha\ep$ starts from zero and becomes back to $zero$ at 
the end of the fast-forward.
%\begin{figure}[h]
%\begin{center}
%\includegraphics[width=7cm]{figure1.eps}% Here is how to import EPS art
%\end{center}
%\caption{\label{fig:epsart} Time dependence of $\ep\alpha$.}
%\label{alpha}
%\end{figure}

\subsection{Fast-forward of adiabatic squeezing of wave packet 
with electro-magnetic field}
\begin{figure}[h]
\begin{center}
\includegraphics[width=7cm]{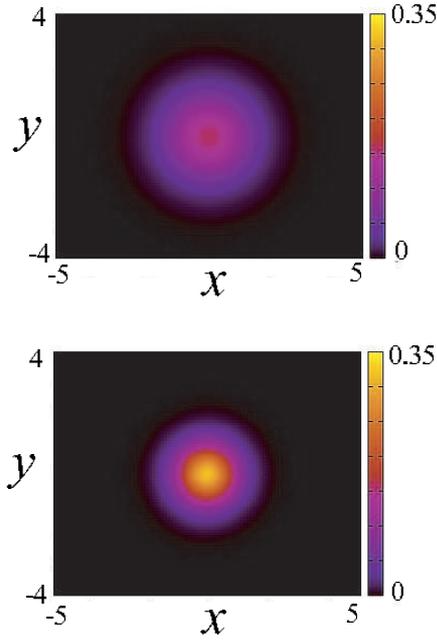}% Here is how to import EPS art
\end{center}
\caption{\label{fig:epsart} Wavefunction $|\Psi_{FF}|$ 
profile before (upper figure) and after (lower figure) the squeezing. }
\label{sq}
\end{figure}

\begin{figure}[h]
\begin{center}
\includegraphics[width=7cm]{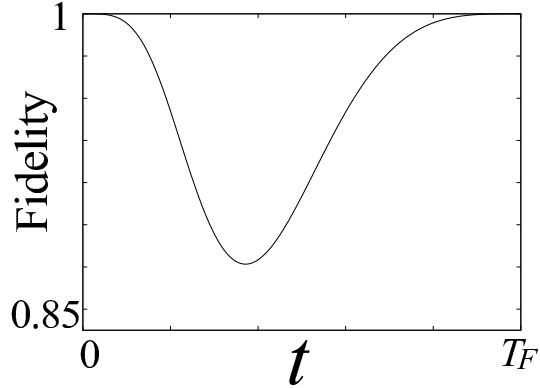}% Here is how to import EPS art
\end{center}
\caption{\label{fig:epsart} Time dependence of fidelity.}
\label{fid_sq}
\end{figure}

We consider an adiabatically squeezed wave packet (WP)
in two dimensions under 
the adiabatically increasing MF  
\begin{eqnarray}
\largeB_0 = (0,0,R(t)),
\end{eqnarray}
where $R(t)=R_0+\ep t$ as given in equation (\ref{R}).
The vector potential corresponding to the MF can be taken as
\begin{eqnarray}
\largeA_0 = (-R(t)y,0,0).
\label{a0101}
\end{eqnarray}
%The potential $V_0$ is put to be zero.
The lowest energy eigenstate $\phi_{n=0}$ 
of the instantaneous Hamiltonian
with energy $E_0=\frac{\hbar R}{2m_0}$
%Then the corresponding $\phi_{n=0}$ in equation (\ref{Psireg}) 
is given by 
\begin{eqnarray}
\phi_0 = \sqrt{\frac{R(t)}{2\pi\hbar}}e^{-\frac{R(t)}{4\hbar}(x^2+y^2)}
e^{i\frac{R(t)xy}{2\hbar}}.
\end{eqnarray}
The corresponding regularized standard state is written with 
the phase of $O(\ep)$ as 
\begin{eqnarray}
\Psi_0^{(reg)}(x,y,t)=\phi_0e^{i\ep\theta}e^{-\frac{i}{\hbar}\int_0^tE_0(R)dt}
=\sqrt{\frac{R(t)}{2\pi\hbar}}e^{-\frac{R(t)}{4\hbar}(x^2+y^2)}
e^{i\frac{R(t)xy}{2\hbar}}e^{-i\frac{1}{2m_0}\int_0^tR(t)dt}e^{i\ep\theta}.
\label{standard1}
\end{eqnarray}
This WF is squeezed when MF is increased.
We accelerate the manipulation which control the width of WP.
Equation (\ref{theta1}) for $\theta$ is rewritten as
\begin{eqnarray}
\nab^2\theta-\frac{R}{\hbar}x\frac{\pa\theta}{\pa x}
-\frac{R}{\hbar}y\frac{\pa\theta}{\pa y}-\frac{m_0}{2\hbar^2}(x^2+y^2)
+\frac{m_0}{\hbar R} = 0.\label{theta4}
\end{eqnarray} 
It can be easily confirmed that 
\begin{eqnarray}
\theta=-\frac{m_0}{4\hbar R}(x^2+y^2)
\label{theta3}
\end{eqnarray} 
satisfies equation (\ref{theta4}).
Equation (\ref{vtilde}) leads to
the regularized standard potential as
\begin{eqnarray}
V_0^{(reg)} = \ep\tilde{V} = -\ep\frac{xy}{2}.
\end{eqnarray} 
With the use of 
equations (\ref{standard1}) and (\ref{theta3}) in equation (\ref{vff4}),
the driving scalar potential is obtained as
\begin{eqnarray}
\frac{V_{FF}}{\hbar} = [\frac{d\alpha}{dt}\ep\frac{m_0}{4\hbar R}-
\alpha^2\ep^2\frac{3m_0}{8\hbar R^2}](x^2+y^2) - \alpha\ep\frac{xy}{2\hbar},
\label{vff101}
\end{eqnarray} 
where we omitted a spatially uniform term because it is concerns only with the
spatially uniform phase. 
The driving EMF can be obtained from 
equations (\ref{eff}) and (\ref{df1}) with the
use of equations (\ref{aff2}), (\ref{vff101}) and (\ref{a0101}).
For numerical calculation the parameters are chosen as $\frac{m_0}{\hbar}=1.0$,
$T_F=1.0$, $\bar{v}=1.0$, $R_0=1.0$. 
The WP profile is shown in figure \ref{sq} at the initial (upper figure) and 
final (lower figure) time of the
fast-forward. It can be seen the WP is squeezed successfully.
To check the accuracy of the acceleration, we evaluated the fidelity which is 
defined by
\begin{eqnarray}
F = |<\Psi_{FF}(t)|\Psi_0(\lam)>|,
\label{fidelity}
\end{eqnarray}
i.e., the overlap between the fast-forwarded state $\Psi_{FF}(t)$
and the corresponding standard one $\Psi_0(\lam)$.
It is unity when $\Psi_{FF}(t) =  \Psi_0(\lam)$.
We confirmed that the fidelity 
first decreases from unity due to the additional phase $f$ 
of the fast-forwarded state,
but at the final time it becomes unity again (see figure \ref{fid_sq}),
which means the exact fast-forward of the adiabatic state besides from the 
spatially uniform phase factor.
%We estimated the fidelity defined by equation (\ref{fidelity}) 
%and confirmed that it becomes back to unity at the 
%end of the fast-forward.

\subsection{Fast-forward of adiabatic transport in electro-magnetic field}
Here we show the fast-forward of adiabatic transport of WF in 2 dimensions
subjected to EMF,
%Suppose a stationary WP in a confining 
%MF and we consider to transport 
%the WP fast 
without leaving any disturbance on the WF
at the end of the transport.
%If we shift the confining MF rapidly to transport the WP fast, 
%the WF is radically affected by EMF, and it
%causes deformation in spatial profile of the WP.
%And it will oscillate in the confining 
%MF after the transportation due to unfavorable momentum transfer.
%However we can realize an ideal transport without causing
%any disturbance at the end of
%the transportation by applying the present theory of
%fast-forward of the adiabatic transport.
%And the difference in the form of the
%driving potential from that for the systems without MF 
%will be shown.
 
WF takes a form 
$\psi(\largex)e^{-\frac{i}{\hbar}E_nt}$, which is stationary except for 
the adiabatic phase,
in the presence of vector potential 
$\largeA(x,y,z)$ at the initial time.
The MF is adiabatically shifted with infinitesimal velocity $\ep$
in $x$-direction.
The shifted vector potential 
$\largeA_0(\largex,t)$ 
are represented with the use of $\largeA$ as
\begin{eqnarray}
%V_0 = U(x -\ep t,y,z),\\
\largeA_0 = \largeA(x -\ep t,y,z).
\end{eqnarray}
%respectively.
%The resultant electric field of $O(\ep)$ is represented as 
%\begin{eqnarray}
%\largeE_0 = -\frac{d\largeA_0}{dt} = -\ep\frac{d\largeA_0}{dx}.
%\end{eqnarray}
%V_0 = U(x -\ep t,y,z),\\
The corresponding regularized
WF is supposed to be given in the form as
\begin{eqnarray}
\Psi_0^{(reg)} = \phi_ne^{-\frac{i}{\hbar}E_nt}e^{i\ep\theta}.
\label{Psireg}
\end{eqnarray}
$\phi_n$ is the $n$-th energy eigenstate of the instantaneous Hamiltonian
written as
\begin{eqnarray}
\phi_n  &=& \psi(x -\ep t,y,z) = \psi(x-R(t),y,z).\label{phin2}
\end{eqnarray}
$R(t)$ which characterizes
the position of WP in $x$ direction is adiabatically changed as
$R(t) = R_0 + \ep t$ with $\ep \ll 1$ and $R_0=0$.
%\begin{eqnarray}
%R(t) = \ep t.
%\end{eqnarray}
Equation (\ref{phin2}) leads to the relation
\begin{eqnarray}
\frac{\pa \phi_n}{\pa R} = -\frac{\pa \phi_n}{\pa x}.
\label{phin3}
\end{eqnarray}
In the same manner as used in the previous example,
%\cite{masuda2},
we can obtain $\theta$ and $\tilde{V}$ as
\begin{eqnarray}
\theta &=& \frac{m_0}{\hbar}x\label{theta2},\nonumber\\
 \tilde{V} &=& -A_x.
\label{vtilde1}
\end{eqnarray}
Therefore the regularized potential $V_0^{(reg)}(x,y,z,t)$ 
in equation (\ref{sp1}) is given as
\begin{eqnarray}
V_0^{(reg)} = -\ep A_x.
\label{V'2}
\end{eqnarray}
With the use of 
equations (\ref{phin3}) and (\ref{theta2}) in equation (\ref{vff4}),
we have the driving scalar potential
\begin{eqnarray}
V_{FF}(x,y,z,t) = - m_0\frac{d\alpha}{dt}\ep x - \alpha\ep A_x,
\label{vff3}
\end{eqnarray}
where we omitted the spatially uniform term because it concerns only on the
spatially uniform phase of WF.
The driving MF is shifted with time and the corresponding 
vector potential is given from equation (\ref{aff2}) as
\begin{eqnarray}
\largeA_{FF}(x,y,z,t) &=& \largeA(x-R(\lam),y,z).
\label{vector1}
\end{eqnarray}
The driving electric field is given by
\begin{eqnarray}
\largeE_{FF}(x,y,z,t) = \ep\alpha (\frac{\pa \largeA_{FF}}{\pa R}) - \nab V_{FF}.
\end{eqnarray}
It is worth noting that since we have derived $\theta$ without giving 
any specific profile on $\phi_n$, 
the formulas of the driving scalar potential in equation (\ref{vff3})
and driving vector potential in equation (\ref{vector1}) 
are independent of the profile of 
the WF that we are going to transport.
%Comparing with the fast-forward without MF,
%there is an extra term ($-\alpha\ep A_x$) in the driving scalar 
%potential as seen
%in equation (\ref{vff3}).
The resultant electric field due to this term and the 
time derivative of $\largeA_{FF}$ can be interpreted as balancing 
with Lorenz force perpendicular to the transport.

%\subsubsection{Transport in uniform magnetic field}
As a concrete example of such accelerated transport,
we consider a case that a  WP trapped in uniform MF $\largeB_0=(0,0,B)$.
%corresponding to the vector potential
We choose the vector potential as
%\begin{eqnarray}
%\largeA = (0,Bx,0),
%\end{eqnarray}
\begin{eqnarray}
\largeA_0 = (0,B(x-\ep t),0),
\label{a010}
\end{eqnarray}
which leads to $\largeB_0$ and electric field 
%In the regularized adiabatic dynamics, 
%the vector potential is shifted as
of $O(\ep)$ in $y$-direction: 
\begin{eqnarray}
\largeE_0 = -\frac{d\largeA_0}{dt} = (0,\ep B,0).
\end{eqnarray}
In this case, $V_0^{(reg)}=0$ as seen from equation (\ref{V'2}).
The WP is adiabatically moved
due to $\theta$ and $\largeE_0$, while the MF does not change.
From equations (\ref{vector1}) and (\ref{a010}), 
it is obvious that we do not have to change MF for
the fast-forward.
%, because
%the shift of $x$ coordinate does not change the vector potential.
An eigenstate with energy $E_0=\frac{\hbar B}{2m_0}$ 
of the instantaneous Hamiltonian
with $R$ is given as 
\begin{eqnarray}
%\psi(x,y)=\sqrt{\frac{B}{2\pi\hbar}}e^{-\frac{B}{4\hbar}((x-R)^2+y^2)}
%e^{i\frac{B(x-R)y}{2\hbar}}
\phi_n(x,y,t)e^{-i\frac{Bt}{2m_0}},
\end{eqnarray}
%Thus the corresponding $\phi_n$ in equation (\ref{Psireg}) is represented as
with
\begin{eqnarray}
\phi_n(x,y,t)=\sqrt{\frac{B}{2\pi\hbar}}e^{-\frac{B}{4\hbar}((x-R(t))^2+y^2)}
e^{-i\frac{B(x-R(t))y}{2\hbar}}.\label{phi1}
\end{eqnarray}
Note that $\phi_n$ in equation (\ref{phi1}) is a stationary state 
with an instantaneous value of $R$.
We transport this state by the driving field.
In this case, the driving scalar potential in equation (\ref{vff3}) is 
represented as
\begin{eqnarray}
V_{FF}(x,y,z,t) =
- m_0\frac{d\alpha}{dt}\ep x,
% + \alpha\ep By.
\label{vff4_2}
\end{eqnarray}
which leads to the driving electric field in $x$-direction.
The driving electric field in $y$-direction given by
\begin{eqnarray}
E_{FF}^{(y)} = -\frac{dA_{FF}^{(y)}}{dt} = \alpha\ep B
\end{eqnarray}
%The second term which does not exist in the case without MF 
is balancing with
Lorenz force in classical picture.
Without this term the path of WP would be bent in $y$-direction.

\begin{figure}[h]
\begin{center}
\includegraphics[width=7cm]{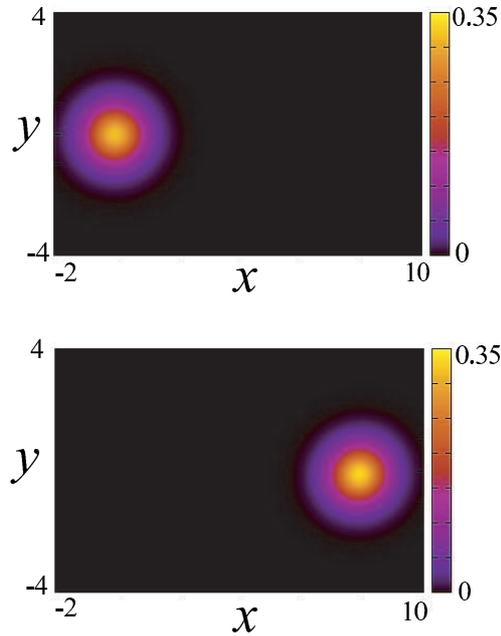}% Here is how to import EPS art
\end{center}
\caption{\label{fig:epsart} Wave packet profile $|\Psi_{FF}|$ at initial 
and final time of the fast-forward.}
\label{trans}
\end{figure}

\begin{figure}[h]
\begin{center}
\includegraphics[width=7cm]{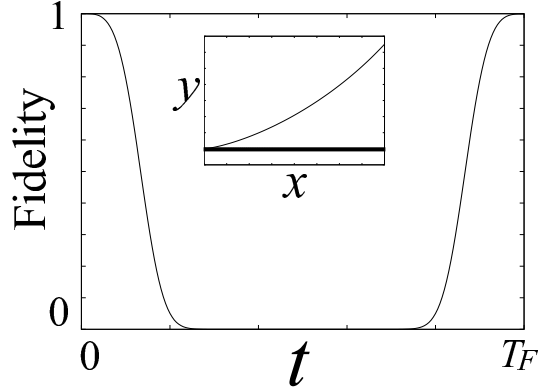}% Here is how to import EPS art
\end{center}
\caption{\label{fig:epsart} Time dependence of fidelity.
The inset represents the path which the wave packet traces under the correct
driving field (thick line) and one without electric field
in $y$-direction (thin line).}
\label{fid_tr}
\end{figure}

In the 
numerical calculation the parameter are chosen as $\frac{m_0}{\hbar}=1.0$, 
$T_{F}=1.0$, $\bar{v}=8.0$ and $R_0=0$.
By applying the driving potential in equation (\ref{vff4_2}),
we accelerate WP. In figure \ref{trans}, the WP profile $|\Psi_{FF}|$
is shown at the initial and final time of the fast-forward.
The WP is transported by distance $8.0$ in time $1.0$ and becomes stationary at
the end.
We confirmed that WP is moved 
without changing its amplitude profile during the 
acceleration. 
%To check the accuracy of the acceleration, we evaluated fidelity which is 
%defined by
%\begin{eqnarray}
%F = |<\Psi_{FF}(t)|\Psi_0(\lam)>|,
%\label{fidelity}
%\end{eqnarray}
%i.e. the overlap between the fast-forwarded state $\Psi_{FF}(t)$
%and the corresponding standard one $\Psi_0(\lam)$.
%It is unity when $\Psi_{FF}(t) =  \Psi_0(\lam)$.
We evaluated the fidelity defined by equation (\ref{fidelity}) 
and confirmed that it becomes back to unity at the 
end of the fast-forward (see figure \ref{fid_tr}).
%The time dependence of the fidelity defined by 
%equation (\ref{fidelity}) is shown in
%figure \ref{fid_tr}.
%We confirmed that the fidelity 
%first decreases from unity due to the additional phase $f$ 
%of the fast-forwarded state,
%but at the final time it becomes unity again.
Thus we have obtained the adiabatically accessible 
target state in a finite time $T_F=1.0$.
The inset in figure \ref{fid_tr} shows the path which the WP 
traces under the correct driving field (thick line) and the path
without electric field in $y$-direction $\alpha\ep B$
%the second term $\alpha\ep By$ in equation (\ref{vff4_2}) 
(thin line). Without this electric field,
the orbit of WP is bent in $y$-direction by MF.

Here we considered the case that WP is trapped by MF.
In \ref{Driving potential in systems with potential} the driving
field is shown in the case that WP is trapped by both MF and 
a scalar potential $V_0$. 

\section{Fast-forward of spin dynamics in magnetic field}
\label{Fast-forward in two-level systems in magnetic field}
We now proceed to the dynamics of spin with $S=\frac{1}{2}$,
by suppressing the orbital degree of freedom.
The fast-forward of spin dynamics in MF is as follows. 
Let us consider the Hamiltonian given by
\begin{eqnarray}
H_0(t) = \frac{1}{2}(B_x\supp(t)\sigma_x+B_y\supp(t)\sigma_y
+B_z\supp(t)\sigma_z),
\label{h0_20}
\end{eqnarray} 
where $B_x\supp$, $B_y\supp$ and $B_z\supp$ denote components of the 
MF and $\sigma_x$, $\sigma_y$, $\sigma_z$ are Pauri matrices. 
In equation (\ref{h0_20}) the negative of Bohr magneton was suppressed for simplicity.
The MF can change with time $t$.
Schr$\ddot{\mbox{o}}$dinger equation can 
be given in $\sigma_z$-diagonal representation as
%be represented 
%with the use of the eigenstates of $H=\sigma_z$ as bases,
\begin{eqnarray}
i\hbar\frac{\pa}{\pa t} 
\left( \begin{array}{cc}
        c_1\supp          \\
        c_2\supp  
\end{array} \right) = 
\left( \begin{array}{cc}
       H_{11}   &  H_{12}             \\
       H_{12}^{\ast}   & -H_{11}               
\end{array} \right)
\left( \begin{array}{cc}
        c_1\supp          \\
        c_2\supp  
\end{array} \right) ,\label{scpsi0}
\end{eqnarray} 
%where the components in the Hamiltonian are related to $B_x\supp$, $B_y\supp$,
%$B_z\supp$ as 
where
\begin{subequations}
\begin{eqnarray}
H_{11}(t) &=& \frac{1}{2}B_z\supp(t)\\
H_{12}(t) &=& \frac{1}{2}(B_x\supp(t)-iB_y\supp(t)).
\end{eqnarray} 
\end{subequations}
The state
\begin{eqnarray}
\Psi_0(t) = 
\left( \begin{array}{cc}
        c_1\supp          \\
        c_2\supp  
\end{array} \right) 
\end{eqnarray} 
is defined as a standard state that we are going to fast-forward.
The exactly fast-forwarded state $\ps(t)$ is defined by $\Psi_0(\lam)$
with $\lam$ in equation (\ref{lam}).
%\begin{eqnarray}
%\ps(t) = \Psi_0(\lam), 
%\end{eqnarray} 
%or equivalently
%\begin{quations}
%\begin{eqnarray} 
%c_1\supa(t) = c_1\supp(\lam),\label{c1a2}\\
%c_2\supa(t) = c_2\supp(\lam),\label{c2a2}
%\end{eqnarray} 
%\end{subequations}
%where $c_1\supa$ and $c_2\supa$ are components of $\ps$ and 
%$\lam$ is defined in equation (\ref{lam}).
Let the Hamiltonian for $\ps$ to be represented as
\begin{eqnarray}
H_\alpha = 
\left( \begin{array}{cc}
       H_{11}\supa   &  H_{12}\supa             \\
       (H_{12}\supa)^{\ast}   & -H_{11}\supa               
\end{array} \right).\label{ha}
\end{eqnarray} 

The definition of the exactly fast-forwarded state in equation (\ref{exact_s}) 
leads to the relation
\begin{eqnarray}
\frac{d\Psi_\alpha}{dt}\Big|_{t=t'}=\alpha(t')
\frac{d\Psi_0}{dt}\Big|_{t=\Lambda(t')}.
\label{dpsi11}
\end{eqnarray} 
%\begin{subequations}
%\begin{eqnarray}
%\frac{dc_1\supa}{dt}\Big|_{t=t'}=\alpha(t')\frac{dc_1\supp}{dt}\Big|_{t=\Lambda(t')},
%\label{dc1a}\\
%\frac{dc_2\supa}{dt}\Big|_{t=t'}=\alpha(t')\frac{dc_2\supp}{dt}\Big|_{t=\Lambda(t')}.
%\label{dc2a}
%\end{eqnarray} 
%\end{subequations}
With the use of Schr$\ddot{\mbox{o}}$dinger equation for $\ps$ with Hamiltonian
$H_\alpha$ together with equations (\ref{scpsi0}) and (\ref{dpsi11}),
%(\ref{c1a2}), (\ref{c2a2}),
%(\ref{dc1a}) and (\ref{dc2a}) 
we have the relation between $H_\alpha$ and $H_0$ as
\begin{eqnarray}
H_\alpha(t)=\alpha(t)H_0(\lam).
\end{eqnarray} 
%\begin{subequations}
%\begin{eqnarray}
%H_{11}\supa(t) = \alpha(t)H_{11}\supp(\lam),\\
%H_{12}\supa(t) = \alpha(t)H_{12}\supp(\lam).
%\end{eqnarray} 
%\end{subequations}
Therefore we can obtain the driving MF, $\largeB_{FF}$, as
\begin{eqnarray}
\largeB_{FF}(t)=
\left( \begin{array}{cc}
        B_x^{FF}(t)\\
        B_y^{FF}(t)\\  
        B_z^{FF}(t)
\end{array} \right) 
=
\left( \begin{array}{cc}
        \alpha(t)B_x\supp(\lam)\\
        \alpha(t)B_y\supp(\lam)\\  
        \alpha(t)B_z\supp(\lam)
\end{array} \right). 
\end{eqnarray}
 
In the fast-forward of spatially distributing WF, 
there was a problem of the anomalous mass reduction.
To resolve this problem
we introduced an additional
phase $f$ on the fast-forwarded state (see equation (\ref{psiff0})), 
but here in spin dynamics we do not have to use it.
%For this fast-forward, 
%no additional phase is necessary,
And we just require to  magnify the MF to 
generate the exactly fast-forwarded state $\ps$.

\subsection{Fast forward of adiabatic spin dynamics}
%So far we showed the regular fast-forward of spin dynamics
%which requires to magnify the MF by $\alpha$ times.
Now, we show the fast-forward of adiabatic spin dynamics.
Suppose that Hamiltonian which is adiabatically changed is represented as
\begin{eqnarray}
H(R(t))=
\left( \begin{array}{cc}
       H_{11}(R(t))   &  H_{12}(R(t))           \\
       H_{12}^{\ast}(R(t))   & -H_{11}(R(t))               
\end{array} \right),\label{ha_ad}
\end{eqnarray}
where $R(t)$ is the parameter which is adiabatically changed as
in equation (\ref{R}), namely, $\ep \ll 1$. The matrix elements
of the Hamiltonian are related to those of MF as,
\begin{subequations}
\begin{eqnarray}
H_{11}(R(t))&=&\frac{1}{2}B_z^{(0)}(R(t))\label{HandB1}\\
H_{12}(R(t))&=&\frac{1}{2}\large{(} B_x^{(0)}(R(t))-iB_y^{(0)}(R(t))\large{)}.
\label{HandB2}
\end{eqnarray}
\label{HandB_al}
\end{subequations}
Suppose 
\begin{eqnarray}
\Psi_0(R(t))=
\left( \begin{array}{cc}
      c_1(R(t))       \\
     c_2(R(t))          
\end{array} \right)e^{-\frac{i}{\hbar}\int_0^tE(R(t))dt}e^{i\Gamma(t)},\label{ha_ad}
\end{eqnarray}
to be an adiabatically evolving state. 
$\Gamma(t)$ is an adiabatic phase, which is common to both component. 
We have
\begin{eqnarray}
E(R)
\left( \begin{array}{cc}
      c_1(R)        \\
     c_2 (R)           
\end{array} \right)
=H(R)\left( \begin{array}{cc}
      c_1 (R)        \\
     c_2(R)            
\end{array} \right),
\label{ha_ad2}
\end{eqnarray}
for an instantaneous Hamiltonian with a parameter $R$.
In this fast-forward, we can not utilize the same manner as used in the 
standard fast-forward of spin dynamics 
because the MF diverges due to infinitely large $\alpha$ ($=O(1/\ep)$).
Thus we need to regularize the standard Hamiltonian.

As we did in the fast-forward of adiabatic orbital dynamics, 
we regularize the system so that the Schr$\ddot{\mbox{o}}$dinger 
equation is fulfilled up to $O(\ep)$ for regularized standard state and 
Hamiltonian.
In this case, however, there is no spatial distribution of WF, and therefore
we do not have to regularize WF with any additional phase
of $O(\ep)$ corresponding to $\theta$ in equation (\ref{reg1}). 
%which corresponds to $\theta$ in equation (\ref{reg1})
%utilized for the fast-forward of spatially distributing WF. 
We simply put a regularized standard state 
without adiabatic phase $\Gamma$ as
\begin{eqnarray}
\Psi_0^{(reg)}(R(t))=
\left( \begin{array}{cc}
      c_1(R(t))        \\
     c_2 (R(t))           
\end{array} \right)e^{-\frac{i}{\hbar}\int_0^tE(R(t))dt},
\label{regs1}
\end{eqnarray}
for which the regularized
Hamiltonian is given by
\begin{eqnarray}
H^{(reg)}_0(R(t))=
\left( \begin{array}{cc}
       H_{11}(R(t)) +\ep h_{11}(R(t))  &  H_{12}(R(t)) +\ep h_{12}(R(t))   \\
       H_{12}^{\ast}(R(t))+\ep h_{12}^{\ast}(R(t))    
& -H_{11}(R(t)) -\ep h_{11}(R(t))         
\end{array} \right),\label{hareg_ad}
\end{eqnarray}
introducing additional terms $h_{11}$ and $h_{12}$.
Schr$\ddot{\mbox{o}}$dinger equation is written as 
\begin{eqnarray}
i\hbar\frac{d\Psi_0^{(reg)}}{dt}=H_0^{(reg)}\Psi_0^{(reg)}.\label{sch1}
\end{eqnarray}
The use of equation (\ref{ha_ad2}) in equation (\ref{sch1}) leads   
\begin{subequations}
\begin{eqnarray}
i\hbar\frac{\pa c_1}{\pa R}=h_{11}c_1+h_{12}c_2, \label{hs1}\\
i\hbar\frac{\pa c_2}{\pa R}=h_{12}^\ast c_1-h_{11}c_2.\label{hs2}
\end{eqnarray}
\label{hss}
\end{subequations}
From equations (\ref{hss}) we can obtain
\begin{subequations}
\begin{eqnarray}
h_{11}=i\hbar (c_1^\ast\frac{\pa c_1}{\pa R}
+c_2\frac{\pa c_2^\ast}{\pa R}), \label{hs1_2}\\
h_{12}=i\hbar (c_2^\ast\frac{\pa c_1}{\pa R}
-c_1\frac{\pa c_2^\ast}{\pa R}), \label{hs2_2}
\end{eqnarray}
\label{hs_al}
\end{subequations}
where $h_{11}$ in equation (\ref{hs1_2}) is purely real because
$\frac{\pa}{\pa R}(|c_1|^2+|c_2|^2)=0$. 

%In the fast-forward of spatially distributing WF, we introduced additional
%phase $f$ on the fast-forwarded state (see equation (\ref{psiff0})), 
%but here we do not have to use it.
The fast-forwarded state is given as 
\begin{eqnarray}
\Psi_{FF}(t)=
\left( \begin{array}{cc}
      c_1(R(\lam))       \\
     c_2(R(\lam))          
\end{array} \right)e^{-\frac{i}{\hbar}\int_0^tE(R(\lam))dt}.\label{psiff_ad1}
\end{eqnarray}
Since there is no additional phase on $\Psi_{FF}$, it is the exactly 
fast-forwarded state except for the adiabatic phase 
common to both components.
We suppose that Hamiltonian $H_{FF}$ drives $\Psi_{FF}$.
The time derivative of equation (\ref{psiff_ad1}) is given by
\begin{eqnarray}
\frac{d\Psi_{FF}}{dt} = {\Big(} \alpha\ep 
  \left( \begin{array}{cc}
\frac{\pa c_1}{\pa R}\\
\frac{\pa c_2}{\pa R}
\end{array} \right) 
-\frac{i}{\hbar}E \left( \begin{array}{cc}
c_1\\
c_2
\end{array} \right) {\Big)} e^{-\frac{i}{\hbar}\int_0^tEdt},
\label{deri3}
\end{eqnarray}
where $\Psi_{FF}(t)$, $\alpha(t)$, $E(R(\lam))$, $c_1(R(\lam))$ and 
$c_2(R(\lam))$ are abbreviated by
$\Psi_{FF}$, $\alpha$, $E$, $c_1$ and 
$c_2$, respectively, and the same abbreviations are used 
hereafter in this section.
Schr$\ddot{\mbox{o}}$dinger equation:
\begin{eqnarray}
i\hbar\frac{d\Psi_{FF}}{dt}=H_{FF}\Psi_{FF}
\end{eqnarray}
and equations (\ref{ha_ad2}), (\ref{hss}) and (\ref{deri3}) lead to 
the driving Hamiltonian as
\begin{eqnarray}
H_{FF}(t)&=&
\frac{1}{2} \left( \begin{array}{cc}
       B_z^{(FF)}  &  
   B_x^{(FF)}-iB_y^{(FF)}\\
       B_x^{(FF)}+iB_y^{(FF)} 
& -B_z^{(FF)}     
\end{array} \right) \nonumber\\
&=& \left( \begin{array}{cc}
       H_{11} +\alpha\ep h_{11}  &  
H_{12} +\alpha\ep h_{12}   \\
       H_{12}^{\ast}+\alpha\ep h_{12}^{\ast}    
& -H_{11} -\alpha\ep h_{11}        
\end{array} \right)\label{haff1}
\end{eqnarray}
where $ H_{11}(R(\lam))$,  $ H_{12}(R(\lam))$, $h_{11}(R(\lam))$
and $h_{12}(R(\lam))$ are abbreviated by 
$ H_{11}$, $ H_{12}$, $h_{11}$
and $h_{12}$, respectively.
The driving MF is obtained 
from equations (\ref{HandB_al}), (\ref{hs_al}) 
and (\ref{haff1}) as
\begin{subequations}
\begin{eqnarray} 
B_x^{(FF)}&=&B_x^{(0)}-
2\hbar\ep\alpha\mbox{Im}(c_2^\ast\frac{\pa c_1}{\pa R}-
c_1\frac{\pa c_2^\ast}{\pa R} ),\\
B_y^{(FF)}&=&B_y^{(0)}-
2\hbar\ep\alpha\mbox{Re}(c_2^\ast\frac{\pa c_1}{\pa R}-
c_1\frac{\pa c_2^\ast}{\pa R} ),\\
B_z^{(FF)}&=&B_z^{(0)}-
2\hbar\ep\alpha\mbox{Im}(c_1^\ast\frac{\pa c_1}{\pa R}+
c_2\frac{\pa c_2^\ast}{\pa R} ),
\end{eqnarray} 
\label{MF_ad1}
\end{subequations}
where $B_i^{(FF)}(t)$ and $B_i^{(0)}(R(\lam))$  
are abbreviated by $B_i^{(FF)}$ and 
$B_i^{(0)}$, respectively, and  $i$ denotes $x,y,z$.
Noting $\ep\alpha=O(1)$, the excess field, $i.e.$, the difference between
$\largeB_{FF}$ and $\largeB_0$ gives a nontrivial contribution.
By applying $\largeB_{FF}=(B_x^{(FF)},B_y^{(FF)},B_z^{(FF)})$ given in 
equation (\ref{MF_ad1}),
we can accelerate adiabatic dynamics and obtain, 
in any short time, the exact target state
except for the common phase between both components.
It is obvious that $\Psi_{FF}$ and $\largeB_{FF}$ coincide with $\Psi_0$
and $\largeB_{0}$, respectively, at the initial and final time.

In closing this subsection, it should be emphasized:
Our purpose in this paper lies in
the fast-forward of the adiabatic dynamics
except for the uniform phase.
Therefore we suppressed the adiabatic phase in equation (\ref{regs1}).
Equation (\ref{MF_ad1}) is a result of such a simplified procedure.
On the other hand, we can also accelerate the adiabatic dynamics
with the adiabatic phase being included in equation (\ref{regs1}).
The resultant driving field is slightly different from equation (\ref{MF_ad1}),
which is described in 
\ref{Fast-forward with adiabatic phase}. 
There, the driving field proved to be equal to the
one obtained recently by Berry (Berry 2009).
%\cite{berry2}.
%However, equation (\ref{MF_ad1}) can lead to the acceleration of the adiabatic dynamics aside from
%the adiabatic phase.

\subsection{Examples}
As an example we consider the adiabatic dynamics in which MF 
is rotated adiabatically
into opposite direction while its magnitude is kept constant.
Let the MF be written as 
\begin{eqnarray}
\largeB=B\largee_r=
B \left( \begin{array}{cc}
    \sin\theta(t)\cos\varphi   \\
    \sin\theta(t)\sin\varphi \\
  \cos\theta(t)
\end{array} \right),\label{MF1}
\end{eqnarray}
where 
\begin{eqnarray}
\theta(t) = R(t).\label{thetaR1}
\end{eqnarray}
$R(t)$ is given by equation (\ref{R}) with $R_0=0$. 
$B$ and $\varphi$ remain constant.
$\largee_r$ is a unit vector pointing the direction of MF.
Hamiltonian for this MF is represented as
\begin{eqnarray}
H(R(t))&=&
\frac{B}{2} \left( \begin{array}{cc}
      \cos(R(t))  &  
   \sin(R(t))e^{-i\varphi}\\
        \sin(R(t))e^{i\varphi}
& -\cos(R(t))  
\end{array} \right). \label{ha_ad3}
\end{eqnarray}
Under this magnetic field, 
the regularized standard state corresponding to 
an adiabatic state with eigenvalue $\lambda_+=\frac{B}{2}$
is given by
\begin{eqnarray}
\Psi_0^{(reg)}&=&
\left( \begin{array}{cc}
      \cos\frac{R(t)}{2}\\
      e^{i\varphi}\sin\frac{R(t)}{2}
\end{array} \right)e^{-\frac{iB}{2\hbar}t}. \label{psireg3}
\end{eqnarray}
From equations (\ref{hs_al}), (\ref{thetaR1}) and (\ref{psireg3}), 
we have 
\begin{subequations}
\begin{eqnarray}
h_{11} &=& 0,\label{h11_3}\\
%h_{12} &=& \frac{\hbar^3B^2}{8}[(\cos^2\frac{R(t)}{2}-\sin^2\frac{R(t)}{2})\sin\varphi
%+i\cos\varphi].\label{h12_3}
h_{12} &=& -\frac{\hbar}{2}(\sin \varphi+i\cos\varphi).\label{h12_3}
\end{eqnarray}
\label{h_3_al}
\end{subequations}
From equations (\ref{haff1}), (\ref{MF_ad1}), and (\ref{h_3_al}), 
we can obtain the driving MF as
\begin{subequations} 
\begin{eqnarray}
%B_x^{(FF)}(t) &=& B\sin\theta(R)\cos\varphi+\frac{\hbar^2B^2}{4}(\cos^2\frac{R}{2}
%-\sin^2\frac{R}{2})\sin\varphi\\
%B_y^{(FF)}(t) &=& B\sin\theta(R)\sin\varphi+\frac{\hbar^2B^2}{4}\cos\varphi\\
%B_z^{(FF)}(t) &=& B\cos(R),
B_x^{(FF)}(t) &=& B\sin(R(\lam))\cos\varphi-\ep\alpha(t)\hbar\sin\varphi,\\
B_y^{(FF)}(t) &=& B\sin(R(\lam))\sin\varphi+\ep\alpha(t)\hbar\cos\varphi,\\
B_z^{(FF)}(t) &=& B\cos(R(\lam)),
\end{eqnarray}
\end{subequations}
where $\lam$ is given by equation (\ref{lam}).
In numerical calculation the parameters are chosen as $\bar{v}=\pi$, $T_F=1.0$
$\varphi=0.0$, $R_0=0.0$, $R(T_F)=\pi$ and $B=1.0$. 
Therefore in the fast-forward, the direction of MF switched into
opposite direction in time $T_F$. 
Initially the state is set as
\begin{eqnarray}
\Psi_{FF}(t=0) =   
\left( \begin{array}{cc}
      1\\
      0
\end{array} \right).
\end{eqnarray}
Spin is rotated by the applied $\largeB_{FF}$ 
and points the opposite direction $\left( \begin{array}{cc}
      0\\
      1
\end{array} \right)$ 
at the end of the 
fast-forward. 
\begin{figure}[h]
\begin{center}
\includegraphics[width=7cm]{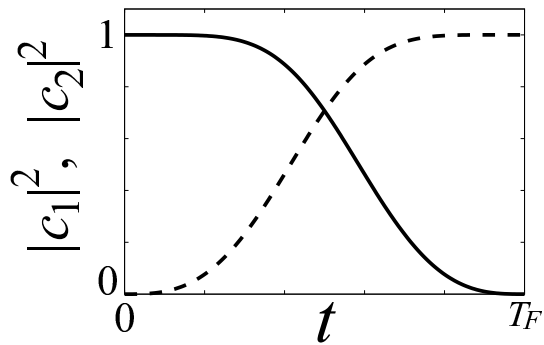}% Here is how to import EPS art
\end{center}
\caption{\label{fig:epsart} The time dependence of 
$|c^{(FF)}_1|^2$ (solid line) and $|c^{(FF)}_2|^2$ (broken line).}
\label{cs}
\end{figure}
In figure \ref{cs}, the dynamics of the spin state: $|c_1^{(FF)}(t)|^2$ and
$|c_2^{(FF)}(t)|^2$ are shown. It is confirmed that spin is flipped exactly
from up to down at the final time $T_F$.
The spin state becomes stationary after the acceleration again.
We also confirmed that the fidelity defined between $\Psi_{FF}(t)$ and
$\Psi_0(\lam)$ is unity during the fast-forward because there is no additional 
phase on the fast-forwarded state.
We can control
the direction of spin more generally, in any desired short time, 
by making $\varphi$
changing through $R$ like $\theta$.
%Here we showed the fast-forward of spin dynamics in MF.
%It is also possible to accelerate the dynamics of two level system in 
%oscillating electromagnetic field
%in a similar way. It is shown in Appendix \ref{FF in Two-Level Systems}.
\\

\begin{figure}[h]
\begin{center}
\includegraphics[width=7cm]{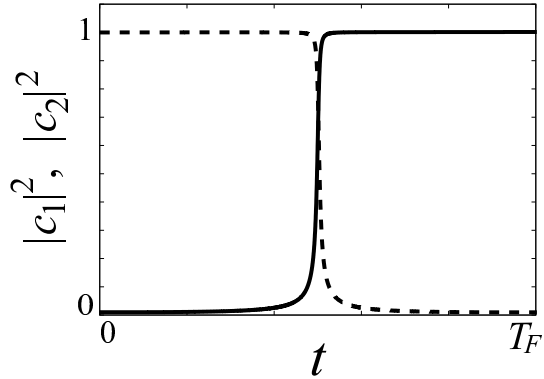}% Here is how to import EPS art
\end{center}
\caption{\label{fig:epsart} The time dependence of 
$|c^{(FF)}_1|^2$ (solid line) and $|c^{(FF)}_2|^2$ (broken line).}
\label{lzm_fig}
\end{figure}

%\subsection{Fast-forward of Landau-Majorana-Zener model}
As another example, we show the fast-forward of adiabatic dynamics
in Landau-Zener (LZ) model 
(Landau 1932; Zener 1932).
%\cite{land,zene}.
We consider MF:
\begin{eqnarray}
\largeB(t)=
\left( \begin{array}{cc}
        \Delta\\
        0\\  
        R(t)
\end{array} \right), 
\end{eqnarray} 
where $\Delta$ is a constant, and $R(t)$ is given in
equation (\ref{R}) with large negative 
constant $R_0$.
The Hamiltonian is given by 
\begin{eqnarray}
H(R(t))  =  \frac{1}{2}
\left( \begin{array}{cc}
        R(t) & \Delta\\
        \Delta & -R(t)
\end{array} \right).
\end{eqnarray} 
The adiabatic state with eigenvalue $\lambda_+=\frac{\sqrt{R^2+\Delta^2}}{2}$
is given by
\begin{eqnarray}
\Psi_0^{(reg)}(t) 
%\left( \begin{array}{cc}
%        c_1      \\
%        c_2  
%\end{array} \right) = 
=\left( \begin{array}{cc}
        -\Delta/s       \\
        \frac{R-\sqrt{R^2+\Delta^2}}{s} 
\end{array} \right)
e^{-\frac{i}{\hbar}\frac{\sqrt{R^2+\Delta^2}}{2}t},
\label{lzm}
\end{eqnarray} 
where 
\begin{eqnarray}
s\equiv \{2\sqrt{R^2+\Delta^2}(\sqrt{R^2+\Delta^2}-R)\}^{1/2}.
\end{eqnarray} 
%with eigen energy $\lambda_+=\sqrt{R^2+\Delta^2}$.
Here we have
\begin{subequations}
\begin{eqnarray}
\frac{\pa c_1}{\pa R}&=&-\frac{1}{2\sqrt{2}}\frac{\Delta}{Q^{5/2}}
(Q-R)^{\frac{1}{2}}
\\
\frac{\pa c_2}{\pa R}&=&
\frac{1}{2\sqrt{2}}\frac{(Q-R)^{1/2}(Q+R)}{Q^{5/2}},
\end{eqnarray} 
\label{dcdrlzm}
\end{subequations} 
where 
\begin{eqnarray}
Q\equiv \sqrt{R^2+\Delta^2}.
\end{eqnarray} 
By using equations (\ref{lzm}) and (\ref{dcdrlzm}) in equation (\ref{MF_ad1}), 
we have the driving field as
\begin{eqnarray}
\largeB_{FF}(t)=
 \left( \begin{array}{cc} \Delta \\
    -\ep\alpha(t)\hbar\frac{\Delta}{R(\lam)^2+\Delta^2}  \\
    R(\lam)
\end{array} \right),\label{lzmB}
\end{eqnarray}
The $y$-component of $\largeB_{FF}$ is identical to the one appearing in 
Landau-Majorana-Zener model 
(Landau 1932; Majorana 1932; Zener 1932; Berry 2009).
%\cite{berry2,land,majo,zene}.
The dynamics of the spin state driven by $\largeB_{FF}$ 
is shown in figure \ref{lzm_fig}
during the fast-forward.
The parameters were chosen as $\bar{v}=100.0$, $T_F=1.0$, $R_0=-50.0$ and $\Delta=1.0$.
We confirmed the fidelity is unity throughout the acceleration.
Interestingly, with the use of the formula different from equation
(\ref{MF_ad1}), Berry also obtained equation (\ref{lzmB}).
This mystery is solved in \ref{Fast-forward with adiabatic phase}.

\section{Conclusion}
\label{Conclusion}
We have presented the theory of the fast-forward 
of quantum adiabatic dynamics in 
electro-magnetic field (EMF).
We derived the driving EMF which accelerates
the adiabatic dynamics and enables to obtain the final adiabatic states 
besides from the spatially uniform phase in
any desired short time, while the final state is 
accessible after infinite time in the adiabatic dynamics.
In the acceleration (fast-forward), 
the initial state is stationary and it becomes back to the
stationary state at the end of the fast-forward without 
leaving any disturbance on the WF.
For the fast-forward of adiabatic orbital dynamics of a charged particle, 
we must control the driving filed, but we do not have to magnify 
the magnetic field from that of adiabatic dynamics,
while in the standard fast-forward, the magnification of the magnetic field
is inevitable.
%We control it as the same intensity as in the adiabatic dynamics.
As typical examples, we showed fast-forward of adiabatic
wave packet squeezing by magnetic field and adiabatic transport in 
EMF.
Furthermore we showed the fast-forward of adiabatic
spin dynamics in time-dependent magnetic field.
%By using this theory, we can point a stationary spin state 
%into any direction.   
The distinction between the present theory and Kato-Berry's transitionless
quantum driving was elucidated.

%\appendix

\begin{acknowledgements}
%S.M. thanks Japan Society for the Promotion 
%of Science for its financial support.
K.N. is grateful for the support received through a project of the Uzbek Academy
of Sciences (FA-F2-084).
We thanks 
D. Matraslov, B. Baizakov, B. Abdullaev, M. Musakhanov,
S. Tanimura and S. Sawada for useful discussions.
\end{acknowledgements}

\appendix{Driving scalar potential in systems with scalar potential $V_0$}
\label{Driving potential in systems with potential}
Here the driving scalar potential of the fast-forward of the 
systems with potential $V_0$ is shown.
The driving vector potential is given by the same form as in the case without 
$V_0$.
%\subsection{Regular fast-forward}

For the fast-forward of regular (non-adiabatic) dynamics with the Hamiltonian
$H_0 = \frac{1}{2m_0}(\largep+\frac{e}{c}\largeA_0)^2+V_0$, we have the 
driving scalar potential $\tilde{V}_{FF}$ as
\begin{eqnarray}
%\frac{V_{FF}}{\hbar} &=& \alpha\frac{V_0}{\hbar} - \frac{df}{dt}
% - (\alpha-1)\frac{\hbar}{2m_0}
%\mbox{Re}[\nab^2\Psi_0/\Psi_0]\nonumber\\
%&&-\frac{\hbar}{m_0}\nab f\cdot\mbox{Im}
%[\nab\Psi_0/\Psi_0] -\frac{\hbar}{2m_0}(\nab f)^2
%-\frac{\hbar}{m_0}\frac{\alpha\largeA_0}{\hbar}\cdot\nab f\nonumber\\
%&&-\frac{\hbar}{2m_0}\alpha(\alpha-1)\frac{\largeA_0^2}{\hbar^2}-
%(\alpha-1)\frac{c_0}{\hbar}|\Psi_0|^2\Psi_0,
\tilde{V}_{FF}(\largex,t)=\alpha(t)V_0(\largex,\lam)+V_{FF}^{(1)}(\largex,t),
\label{eqa1}
\end{eqnarray}
where $V_{FF}^{(1)}$ is defined by $V_{FF}$ in equation (\ref{vff05}).
%where $f(\largex,t)$, $\Psi_0(\largex,\lam)$, $\alpha(t)$, 
%$\largeA_0(\largex,\lam)$, $V_0(\largex,\lam)$ and
%$V_{FF}(\largex,t)$ are abbreviated by $f$, $\Psi_0$, $\alpha$, $\largeA_0$,
%$V_0$ and $V_{FF}$, respectively.
The driving vector potential is given by the same form as in 
equation (\ref{bff1}).

%\subsection{Fast-forward of adiabatic dynamics}
The driving scalar 
potential for the fast-forward of adiabatic dynamics with Hamiltonian
\begin{eqnarray}
H_0 = \frac{(\largep+\largeA_0(\largex,R(t)))^2}{2m_0} + V_0(\largex,R(t)),
\end{eqnarray}
is given by
\begin{eqnarray}
%\frac{V_{FF}}{\hbar} &=& \frac{V_0}{\hbar}
%-\frac{d\alpha}{dt}\ep\theta-\alpha^2\ep^2\frac{\pa\theta}{\pa R}
%-\frac{\hbar}{2m_0}\alpha^2\ep^2(\nab\theta)^2
%-\frac{\hbar}{m_0}\alpha\ep\frac{\largeA_0}{\hbar}\cdot\nab\theta
%\nonumber\\
%&&-\alpha\ep\mbox{Im}[\frac{\pa\phi_n}{\pa R}/\phi_n]
%-\alpha\ep\frac{\hbar}{m_0}
%\mbox{Im}[\frac{\nab\phi_n}{\phi_n}]\cdot\nab\theta,
\tilde{V}_{FF}(\largex,t)=V_0(\largex,\lam)+V_{FF}^{(2)}(\largex,t),
\end{eqnarray}
where $V_{FF}^{(2)}$ is defined by $V_{FF}$ in equation (\ref{vff4}).
%where $V_{FF}(\largex,t)$, $\alpha(t)$,
%$\phi_n(\largex,R(\lam))$, $\largeA_0(\largex,R(\lam))$ and 
%$\theta(\largex,R(\lam))$ 
%are
%abbreviated by $V_{FF}$, $\alpha$, $\phi_n$, $\largeA_0$
%and $\theta$ respectively.

%\subsection{Fast-forward of adiabatic transport}
In section \ref{Examples}, we showed the fast-forward of adiabatic transport
of wave packet trapped by magnetic field.
Here we show the driving scalar potential for adiabatic transport of wave packet
trapped by electric and magnetic  
field with $V_0$.
In adiabatic dynamics, the trapping scalar potential is also shifted as
\begin{eqnarray}
V_0 = U(x -\ep t,y,z),
\end{eqnarray}
as well as vector potential, where $U$ is a static trapping scalar potential.
In such case, the driving scalar potential is represented as
\begin{eqnarray}
\tilde{V}_{FF}(\largex,t) = U(x -R(\lam),y,z) + V_{FF}^{(3)}(\largex,t),
%- m_0\frac{d\alpha}{dt}\ep x - \alpha\ep A_x.
\end{eqnarray} 
where $V_{FF}^{(3)}$ is defined by $V_{FF}$ in equation (\ref{vff3}).

\appendix{Fast-forward of adiabatic spin dynamics with adiabatic phase}
\label{Fast-forward with adiabatic phase}
%We represent the driving field in equation (\ref{MF_ad1}) by standard magnetic
%filed.
We compare the driving field 
in equation (\ref{MF_ad1}) with the one obtained by Berry 
(Berry 2009)
%\cite{berry2}
with the use of Kato's formalism 
(Kato 1950).
%\cite{kato}.
Under the magnetic field
\begin{eqnarray}
\largeB_0(R(t))&=&
B_0\left( \begin{array}{cc}
   \sin\theta\cos\varphi\\  
   \sin\theta\sin\varphi\\
   \cos\theta
\end{array} \right),
\label{eq3_2}
\end{eqnarray}
$c_1$ and $c_2$ in equation (\ref{ha_ad}) are represented as
\begin{subequations}
\begin{eqnarray}
c_1 &=& \cos\frac{\theta}{2}\\
c_2 &=& e^{i\varphi}\sin\frac{\theta}{2},
\end{eqnarray}
\label{c1c2_1}
\end{subequations}
respectively.
Here all of $\largeB_0$, $\theta$ and $\varphi$ are dependent on $R(t)$
which is slowly changing in time.
Substituting equation (\ref{c1c2_1}) into equation (\ref{MF_ad1}), 
our driving field is expressed as
\begin{subequations}
\begin{eqnarray}
B_x^{(FF)}(t) &=& B_x^{(0)}(R) - 2\hbar\ep\alpha (\frac{\pa}{\pa R}
\htheta\sin\varphi +
\frac{\pa}{\pa R}\varphi\sin\htheta\cos\htheta\cos\varphi),\\
B_y^{(FF)}(t) &=& B_y^{(0)}(R) + 2\hbar\ep\alpha (\frac{\pa}{\pa R}
\htheta\cos\varphi -
\frac{\pa}{\pa R}\varphi\sin\htheta\cos\htheta\sin\varphi),\\
B_z^{(FF)}(t) &=& B_z^{(0)}(R) + 2\hbar\ep\alpha\frac{\pa}{\pa R}
\varphi\sin^2\htheta.
\end{eqnarray}
\label{bff412_2}
\end{subequations}

So far,
in our main text,
we derived the driving field which accelerates the adiabatic dynamics
except for the uniform phase. Now, let us choose the regularized standard state 
with the adiabatic phase as
\begin{eqnarray}
\Psi_0^{(reg)}(R(t))=
\left( \begin{array}{cc}
      c_1(R(t))        \\
     c_2 (R(t))           
\end{array} \right)e^{-\frac{i}{\hbar}\int_0^tE(R(t))dt}e^{i\xi(t)},
\end{eqnarray}
where $\xi(t)$ is the adiabatic phase given by
\begin{eqnarray}
\xi(t) &=& i\int_0^tdt'
(c_1^\ast\frac{\pa c_1}{\pa t} + c_2^\ast\frac{\pa c_2}{\pa t}),\\
&=& i\ep\int_0^tdt'
(c_1^\ast\frac{\pa c_1}{\pa R} + c_2^\ast\frac{\pa c_2}{\pa R}).
\label{xi_3}
\end{eqnarray}
In this case, $h_{11}$ and $h_{12}$ in the regularized Hamiltonian 
corresponding to
equation (\ref{hareg_ad}) are represented as
\begin{subequations}
\begin{eqnarray}
%h_{11}&=&i\hbar (c_1^\ast\frac{\pa c_1}{\pa R}
%+c_2\frac{\pa c_2^\ast}{\pa R}) + i\xi(|c_1|^2-|c_2|^2),\\
%h_{12}&=&i\hbar (c_2^\ast\frac{\pa c_1}{\pa R}
%-c_1\frac{\pa c_2^\ast}{\pa R})+2i\xi c_1c_2^\ast, 
h_{11}&=&i\hbar (c_1^\ast\frac{\pa c_1}{\pa R}
+c_2\frac{\pa c_2^\ast}{\pa R}) + L(|c_1|^2-|c_2|^2),\\
h_{12}&=&i\hbar (c_2^\ast\frac{\pa c_1}{\pa R}
-c_1\frac{\pa c_2^\ast}{\pa R})+2L c_1c_2^\ast, 
\end{eqnarray}
\label{hs_al_3}
\end{subequations}
where
\begin{eqnarray}
L\equiv -i\hbar(c_1^\ast\frac{\pa c_1}{\pa R} + c_2^\ast\frac{\pa c_2}{\pa R}).
\end{eqnarray}
In the analogous way used in Section 
\ref{Fast-forward in two-level systems in magnetic field},
the driving Hamiltonian is represented as equation (\ref{haff1}).
With the use of equations (\ref{haff1}), (\ref{c1c2_1}), 
(\ref{xi_3}) and (\ref{hs_al_3}), 
we obtain the driving field as 
\begin{subequations}
\begin{eqnarray}
B_x^{(FF)}(t) &=& B_x^{(0)}(R) - \hbar\ep\alpha(t) (\frac{\pa\theta}{\pa R}
\sin\varphi + \frac{\pa\varphi}{\pa R}\sin\theta\cos\theta\cos\varphi),\\
B_y^{(FF)}(t) &=& B_y^{(0)}(R) + \hbar\ep\alpha (\frac{\pa\theta}{\pa R}
\cos\varphi -
\frac{\pa\varphi}{\pa R}\sin\theta\cos\theta\sin\varphi),\\
B_z^{(FF)}(t) &=& B_z^{(0)}(R) + \hbar\ep\alpha\frac{\pa\varphi}{\pa R}
\sin^2\theta.
\end{eqnarray}
\label{bff412}
\end{subequations}
This kind of driving field was already 
obtained by Berry (Berry 2009),
%\cite{berry2}, 
but is different 
from ours which accelerates the adiabatic 
dynamics without the adiabatic phase. 
In fact, we see the polar angle $\theta$ in equation (\ref{bff412}) 
wherever $\theta /2$
appears in equation (\ref{bff412_2}).
Interestingly, in our examples of spin inversion and LZ model
where no adiabatic phase appears (:$\frac{\pa \varphi}{\pa R}=0$),
both equations (\ref{bff412_2}) and (\ref{bff412}) 
lead to the identical driving field.
In particular the $y$-component of $\largeB_{FF}$ in Landau-Majorana-Zener 
model 
(Landau 1932; Majorana 1932; Zener 1932)
%\cite{land,majo,zene} 
is available from both equations (\ref{bff412_2}) and (\ref{bff412}).
%and 
%equivalent to the one derived by Berry \cite{berry2}.
The distinction between equations (\ref{bff412_2}) and (\ref{bff412})
will manifest itself in the case of the winding Landau-Zener model
(Berry 1990; Nakamura \& Rice 1994; Bouwmeester $et. al.$ 1996),
where the adiabatic phase is non-vanishing.

\label{lastpage}
\end{document}